\documentclass[a4paper,11pt]{article}
\pdfoutput=1  

\usepackage{jheppub} 

\usepackage[T1]{fontenc} 

\usepackage{amsfonts}

\title{\boldmath Transgression branes in seven dimensions }

\author[a,1]{Pablo Mora,\note{Corresponding author.}}
 
\affiliation[a]{Departamento de Ciencias F\'{\i}sicas, 
Centro Universitario Regional Este (CURE), Universidad de la Rep\'ublica, Uruguay, Ruta 9 km. 207, Rocha, Uruguay}
 
\emailAdd{pablomora@cure.edu.uy}

\abstract{A seven dimensional theory with branes, described by a Transgression Action is investigated. The theory has special properties 
owing to the factorization of the trace used to define the Transgression form. 
It is shown that in some cases the theory is similar to the models of branes proposed by Nepomechie, Wu and Zee as a higher dimensional generalization of models of particles with fractional statistics of Wilczek, Wu and Zee. Static Euclidean solutions are presented, and compared with similar solutions of Nepomechie,Wu and Zee. Then Lorentzian static solutions are considered, based in solutions of constant curvature black holes presented by Ba\~nados and further studied by Guilleminot, Olea and Petrov. Topological phases and their implications for quantum statistics are discussed.}

\begin{document} 
\maketitle
\flushbottom

\section{Introduction}

Chern-Simons branes were proposed  in Ref.\cite{Mora-Nishino20001} as models of extended objects in the framework of Chern-Simons (super)gravity theories \cite{Achucarro:1987vz,Chamseddine:1990gk, Chamseddine:1989nu, MuellerHoissen:1990vf, Banados:1993ur, Banados:1996hi, Zanelli:2012px}. Some properties of this models were discussed in my previous works  \cite{Mora:2021wmz,Mora:2023ngp}, were the pertinent references were given \cite{Dixon:1991xz,Green:1989nn,Troncoso1997va, Troncoso1998ng,Horava:1997dd}, as well as references to other papers on Chern-Simons branes \cite{Edelstein:2008ry,Miskovic:2009dd,Edelstein:2010sx,Edelstein:2010sh,Edelstein:2011vu,Kastikainen:2020auf,Frey:2019fqz,Ertem:2012qv}.

One novel property discussed in ref.\cite{Mora:2023ngp} was the enhanced gauge invariance of some of this models, owing the the factorization of the invariant traces considered\footnote{One can certainly write Actions with Lagrangians given as products of traces, but what the factorization property implies is that one cannot add terms with the same symmetry and traces not factorized, 
because those terms do not exist. If they do, they should be added, in accord with the so called "totalitarian principle", invoked for instance by Murray Gell-Mann: "what is not forbidden is mandatory". }, and the so called Inflow Mechanism.  The enhanced invariance discussed in \cite{Mora:2023ngp} would work in several dimensions and for different symmetry groups\footnote{A recent reference discussing symmetry groups with the sort of factorization property required is \cite{Montero:2025ayi}.}. In this work I will focus on the case of a seven dimensional model with 2-branes, with gauge group $SO(8)$ (Euclidean) or $SO(6,2)$ (Lorentzian), mentioned in \cite{Mora:2023ngp}, which has a rich set of properties.

The model considered here turns out to be connected with the higher dimensional branes considered by 
Nepomechie, Wu and Zee in refs.\cite{Nepomechie1984,Wu:1988py} and further investigated by Tze and Nam in ref.\cite{Tze:1989jv}, which are a generalization of 
theories in 2+1 dimensions with particles with fractional statistics 
(anyons)\cite{Wilczek:1981du,Wilczek:1982wy,Wilczek:1983cy,Wu:1984kd,Jain:1989tx}. 
The 2+1 dimensional models are the foundation of an effective Chern-Simons field theory of the Quantum Hall Effect (QHE), introduced in 
refs.\cite{Zhang:1988wy,Zhang:1992eu}, and then derived and studied from general principles in 
refs.\cite{Frohlich:1991wb,Zee:1995avy,Zee:2003mt}.

There are generalizations of the QHE in higher dimensions introduced by Zhang and Hu \cite{Zhang:2001xs}, with an effective field theory description \cite{Bernevig:2002eq} similar to the one in 2+1 dimensions. Further generalizations and studies of these models were done
in refs.\cite{Bernevig:2003yz,Karabali:2002im,Karabali:2003bt,Karabali:2004km,Agarwal:2024fgp,Karabali:2026non,Hasebe:2014nia,Hasebe:2016tjg}.

Zhang Shou-Cheng, co-proponent of the higher dimensional QHE, suggested in \cite{Zhang_2004} that one could understand Fundamental Physics (including High Energy Physics) in this framework, as some long distance or low energy limit of 
some unknown possibly much more complicated truly microscopic dynamic and structure, as it often happens in Condensed Matter Physics. All the physics we see would consist of the lowest energy perturbations of some vacuum, or what in High Energy Physics is known as "massless modes" and in Condensed Matter Physics as "gapless excitations"\footnote{The Standard Model does in fact consist of a bunch of massless fields that obtain their effective masses from their interaction with the Higgs field. It is also conformally invariant, except for the Higgs terms. The dimensionful parameters of the Higgs field, and the beyond the Standard Model neutrino masses (whose nature is still unknown) will probably get a deeper explanation in the future.}. A similar point of view has long been advocated also by Wen Xiao-Gang, for instance in his book \cite{Wen:2004ym}.

An alternative vision of Physics, advocated for instance by Einstein \cite{einstein1933method} and Weinberg \cite{Weinberg}, 
sees it as an uncovering of increasingly simpler structures, guided by general principles, that would end with an all encompassing
Unified Theory, which would in retrospect seem unavoidable.

I do not know which of these two visions, if any, is true, but it is remarkable that in both of them a very 
simple formulation of Natural Law is expected. I think there is a chance that the kind of model discussed 
in this work may be relevant in Fundamental Physics, and it is interesting 
that one could motivate them from both of the modes of thinking mentioned before.

The plan of this work is the following:

In Section 2 I review the previous work by my collaborators and myself on Transgression Forms as a framework to improve Chern-Simons
field theories and to formulate theories with Chern-Simons actions and dynamical extended objects, which constitute the basis of the present work.

In Section 3 I review the works of Wilczek, Nepomechie, Wu and Zee \cite{Wilczek:1981du,Wilczek:1982wy,Wilczek:1983cy,Nepomechie1984,Wu:1988py} on particles with fractional statistics (anyons), their Chern-Simons interactions and their generalization to higher dimensions and branes, as well as the very important related works on the Quantum Hall Effect (QHE) in 
2+1 dimensions \cite{Zhang:1988wy,Zhang:1992eu,Frohlich:1991wb,Zee:1995avy,Zee:2003mt} 
and its higher dimensional generalizations 
\cite{Zhang:2001xs,Bernevig:2002eq,Bernevig:2003yz,Karabali:2002im,Karabali:2003bt,Karabali:2004km,Agarwal:2024fgp,Karabali:2026non,Hasebe:2014nia,Hasebe:2016tjg}. 

In Section 4 I look at the relations between the models reviewed in Section 2 and the Transgression Branes discussed in my previous works,
at the level of both actions and solutions to the field equations (for the Euclidean version of the theory). The possibility that the so called Exotic Spheres \cite{Milnor} may be solutions is discussed.

In Section 5 I first review the Constant Curvature Black Hole (CCBH) solutions introduced by Ba\~{n}ados \cite{Banados:1997df} and further studied in Ref.\cite{Guilleminot:2017hfy}, and then show that some of these configurations are Lorentzian solutions to the Transgression Branes action. 

The solutions considered in Sections 4 and 5 satisfy two conditions: i. have vanishing gauge curvature $F=0$ and ii. are singular at the branes. 
It is worthwhile to note that the condition $F=0$ is sufficient but not necessary for a solution, 
as even the condition $STr[F T^A]=0$, with $T^A$ a generator of the group, in seven dimensions and for the groups considered does not automatically imply $F=0$, although the reciprocal is true. The generic equations of motion imply more complicate non linear interactions between the branes and the dynamical background \cite{Mora-Nishino20001,Mora:2021wmz,Mora:2023ngp}, 
with potentially generic more complicated non static solutions. 

Section 6 deals with topological quantum phases and quantum statistics, and the possibility of branes with fractional statistics, 
as a higher dimensional version of anyons.

Finally, Section 7 are the Conclusions, and the Appendices deal with Dirac's Magnetic Monopole \cite{Dirac:1931kp,Dirac:1948um}
and the Belavin-Polyakov-Schwartz-Tyupkin (BPST) Instanton \cite{Belavin:1975fg}
in the so called "Stereographic Representation", and using Complex Numbers (see for instance Ch.0 of Ref.\cite{Naber:1997yu}) 
and Quaternions\footnote{It is interesting to note that Quaternions have a central role in many of the themes 
that come together in this work, namely Hopf fibrations, BPST instantons and Exotic Spheres. In the Commentary 
to his paper [[55b]] in pages 21-23 of the book \cite{Yang:2005yrp} Yang Chen-Ning discussed his (yet?) unsuccessful 
attempts to generalize gauge theories using quaternions, which he hoped would be eventually vindicated, 
as the quaternions are a beautiful mathematical structure.} 
respectively \cite{atiyah1979geometry}. 
These representations are relevant and to be contrasted with the "spherical" description of the same objects in the main text, corresponding to the Wu-Yang magnetic monopole \cite{Wu:1975es}, as described by Zee in Ch.IV.4 of \cite{Zee:2003mt}, and the analogous BPST instanton and anti-instanton, as described in Subsection 4.4 of this work.

\section{Review of Chern-Simons and Transgression theories}

\subsection{Characteristic Classes, Chern-Simons forms and Transgressions}

The content of this Subsection is based on Refs.\cite{Alvarez-Gaume:1984zlq,MOTZ-2006} and references therein.

I will use the notation $STr[something]\equiv <something>$ for the symmetrized trace, or contraction with the symmetric invariant tensor considered. That is, if $T^A$ are the generators of the group considered, then $<T^{A_1}...T^{A_{n+1}}>=g _{A_1...A_{n+1}}$ is an invariant tensor. One important property is that $d<something>=<D(something)>$, where the covariant derivative is $D\equiv d +[A,\; ]$.

The expression $P(F)=<F^{n+1}>$ is an "Invariant polynomial". The invariant polynomial $P(F)$ is closed, $d P(F)=0$, 
because $d P(F)=d<F^{n+1}>=<D(F^{n+1})>   =(n+1)<(DF)F^n>=0$, which follows from the Bianchi identity $DF=0$. 
That means that it is locally exact $<F^{n+1}>=d\mathcal{C}_{2n+1}(A,F)$, where 
$\mathcal{C}_{2n+1}(A,F)$ is the Chern-Simons form
$$\mathcal{C}_{2n+1}(A,F)=(n+1)\int _0^1 ds <AF_s^n>,$$
where $A_s=s A$ and $F_s=dA_s+A_s^2=s dA+s^2 A^2$.

The previous locally valid expressions can be generalized to globally valid 
ones by considering two gauge configurations $A$ and $\overline{A}$. 
Then we have $<F^{n+1}>-<\overline{F}^{n+1}>=d\mathcal{T}_{2n+1}$, with the Transgression form given by
$$\mathcal{T}_{2n+1}=(n+1)\int _0^1 ds <(\Delta A) F_s^n>,$$
where $A_s =sA+ (1-s)\overline{A}$, so that $A_1=A$ and $A_0=\overline{A}$, $\Delta A=A-\overline{A}$ 
and $F_s=dA_s+A_s^2=s dA+s^2 A^2$.

The variation of the Transgression under variations $\delta A$ and $\delta \overline{A}$ is
$$\delta \mathcal{T}_{2n+1}=(n+1)<F^n\delta A>-(n+1)<\overline{F}^n\delta \overline{A}>-n(n+1)d[\int _0^1<\Delta AF_s^{n-1}\delta A_s>].$$
It can be shown that the Transgression form is the difference of the Chern-Simons form for each gauge field plus a total derivative
$$\mathcal{T}_{2n+1}(A,\overline{A})=\mathcal{C}_{2n+1}(A)-\mathcal{C}_{2n+1}(\overline{A})-d C_{2n},$$
with
$$C_{2n}=n(n+1)\int _0^1ds\int_0^1 dt \; t <(\Delta A) A_s F_{st}^{n-1}>.$$
Here $A_s= sA +(1-s)\overline{A}$ and $F_{st}=t dA_s+t^2 A_s^2=t F_s+(t^2-t)A_s^2$.

\subsection{Chern-Simons and Transgression Actions}

Chern-Simons field theories are models in diverse dimensions with the Chern-Simons form for some group as Lagrangian. 
They are relevant both in Condensed Matter Physics \cite{Zhang:1988wy,Zhang:1992eu,Frohlich:1991wb,Zee:1995avy} 
and Fundamental Physics 
\cite{Achucarro:1987vz,Chamseddine:1989nu,Chamseddine:1990gk,MuellerHoissen:1990vf,Banados:1993ur,Banados:1996hi} 
(see \cite{Zanelli:2012px} for a review and detailed references).

In a series of papers by my collaborators Olea, Troncoso and Zanelli, and myself \cite{MOTZ-2004,MOTZ-2006,Mora20141} we extended these 
theories by using Transgression forms (that are truly gauge invariant) as Lagrangians, instead of Chern-Simons forms (that are quasi-invariant), and were able to 
show that a wealth of problems with the conserved charges and black hole thermodynamics are solved at one stroke by this generalization. 
As Steve Jobs once said: "If you do the right things on the top line, the bottom line will follow". 
This pattern has been verified in ulterior works by other researchers, 
as in recent work by Cai, Li and Zhao \cite{Cai:2024tyv}, and references therein.

Chern-Simons branes were introduced in \cite{Mora-Nishino20001} as models of 
extended objects and sources in backgrounds described by Chern-Simons or Transgression theories. 
The Action is given by the sum of integrals of Transgression forms of diverse dimensions for some group, with the higher dimension
being the dimension of the space-time. Some attractive properties of these models were discussed in \cite{Mora:2021wmz,Mora:2023ngp}, 
including that they are fully gauge invariant and generally covariant. The branes are dynamical, as is the background, and can have boundaries. I think that these models fit very well in
Yang's vision \cite{Yang,Yang:2007upv,Yang:2012bim} of the main themes of fundamental physics of the last century. 

As in Refs.\cite{Mora-Nishino20001,Mora:2021wmz,Mora:2023ngp} the dynamical variables 
are the gauge fields given $A=A_{\mu}^A(x)T^Adx^{\mu}$ valued on the gauge group 
$\mathcal{G}$ algebra with generators $T^A$, with $\mu =0,...,D$, with $D$ even, the embedding coordinates of each brane 
$X_{(d+1)}^{\mu}(\xi _{(d+1)}^i)$, where $\xi _{(d+1)}^i$, 
with $i=0,...,d$, with $d$ even, are the coordinates of the $d$ brane, and the intrinsic metrics defined at the boundaries of the branes 
$\gamma _{(d)rs}$, with $r,s=0,...,d-1$. It was argued in \cite{Mora:2021wmz} that only the two-dimensional metric $\gamma _{(2)rs}$ does appear, as it is a non dynamical auxiliary variable. The space-time coordinates are $\xi _{(D+1)}^{\mu}\equiv x^{\mu}$. The branes 
may have boundaries, but those boundaries have no boundary. 
The action is given as a sum of terms of diverse dimensions corresponding to the various branes and 
the space-time in which these branes are embedded
\begin{equation}
S=\sum _{n=1}^{\frac{D}{2}}S_{2n+1}^{(Trans)} +S_{2}^{(Kin)},
\end{equation}
where the Transgression action part is given by
\begin{equation}
S_{2 n+1}^{(Trans)}=k _{2 n+1}\int _{\mathcal{M}^{2n+1}} \mathcal{T}_{2 n+1},
\end{equation}
while the Kinetic part of the action has support on the boundaries of the 
2-branes, if such boundaries exist, and is of the form
\begin{equation}
S_{2}^{(Kin)}=\pm\frac{k_2}{2}\int _{\mathcal{M}^2\equiv \partial \mathcal{M}^{3}}d^2\xi _{(2)}
\sqrt{-\gamma } ~\gamma ^{ij}<\Delta A_i\Delta A_j>,
\end{equation}
where $\Delta A_i=\Delta A_{\mu}^A\frac{\partial X_{(2)}^{\mu}}{\partial \xi _{(2)}^i}T^A$.

In Ref.\cite{Mora:2023ngp} it was shown that if the invariant trace used to define the Transgression factorizes the model has an enhanced gauge invariance, as a consequence of the inflow mechanism. One example of this property is the group $SO(8)$, or its "Lorentzian version"
$SO(6,2)$, which correspond to the models considered in this work. In this case, from eq.(7.4) in \cite{Mora:2023ngp}, we have for the invariant polynomial $Tr F^4=\frac{1}{12}(Tr F^2)^2$ if $n=8$, where the traces are taken in the adjoint representation. This invariant polynomial is an 8-form, suitable to define a Transgression theory in seven dimensions with 2-branes.

\subsection{Orthogonal and de Sitter gauge groups }

The algebra of generators of $SO(D+1)$ with any signature, codified in the metric $\eta _{AB}$, is

$$[J_{AB},J_{CD}]=-J_{AC}\eta _{BD}+J_{AD} \eta _{BC}+J _{BC}\eta _{AD}- J _{BD}\eta _{AC}$$

with $A,B=1,.....,D+1$. The gauge potential or connection for these gauge groups, that allows to define the covariant derivative $D\equiv d +A$,  is

$$A=\frac{1}{2}A^{AB}J_{AB}=\frac{\omega ^{ab}}{2}J_{ab}+\frac{e^a}{l}P_a$$

with $a,b=1,.....,D$, $J_{a, D+1}=P_a$, $\omega ^{ab}$ the spin connection, $e^a$ the vielbein, and $l$ a constant with the dimensions of length.  From its definition $A$ has dimensions of inverse length or mass, that is $[A]=L^{-1}$. The generators are dimensionless $[J]=[P]=1$, as is the vielbein, like the metric. 
Therefore $[\omega ]=L^{-1}$, and we need to introduce an otherwise arbitrary parameter with the dimension of lenght $l$ 
so that $[e/l]=L^{-1}$. The field strength or curvature is

$$F=dA+A^2 =\frac{1}{2}F^{AB}J_{AB}=\frac{\tilde{R}^{ab}}{2}J_{ab}+\frac{T^a}{l}P_a$$
 
where $F^{AB}=dA^{AB}+A^{AC}A_C^{\;B}$, $\tilde{R}^{ab}=R^{ab}\pm \frac{1}{l^2}e^a e^b$ with the sign determined by $\eta _{D+1,D+1}=\mp 1$, 
and  $R^{ab}=\omega ^{ac}\omega _c^{\:b}$ is the standard Riemannian curvature. Also $F^{a,D+1}=\frac{T^a}{l}=\frac{1}{l}[de^a +\omega ^{ab}e_b]$
The dimensions of the field strength an curvature are $[F]=[R]=L^{-2}$, and the dimension of the torsion is $[T]=L^{-1}$.

We will need the symmetrized traces

$$STr[J_{AB}J^{CD}]=\frac{1}{2}[\delta _A^C \delta _B^D-\delta _A^D\delta _B^C]$$

and therefore

$$STr[F^2]=\frac{1}{4}F^{AB}F_{AB}=\frac{1}{4}[\tilde{R}^{ab}\tilde{R}_{ab}+\frac{2}{l^2}T^aT_a].$$

We will consider particular solutions of our Lagrangian such that $F=0$. For these solutions $T^a=0$ and $R^{ab}=\mp \frac{1}{l^2}e^a e^b$ if $\eta _{D+1,D+1}=\mp 1$.
If we take the limit $l\rightarrow\infty$, or more precisely if we look at configurations such that $L/l\ll 1$, where $L$ is the scale characteristic of that configuration, we see that $R^{ab}\rightarrow 0$.

\section{Review of Nepomechie-Wilczek-Wu-Zee models of  particles and branes}

\subsection{The Quantum Hall Effect as a source of insight}

The Quantum Hall Effect (QHE) in its many incarnations has turned out to be a surprisingly rich field, with many very important and interesting experimental and theoretical developments.
To our purposes, the formulation of the effective theory of QHE in 2+1 dimensions the by Zhang, Hansson and Kivelson \cite{Zhang:1988wy}, as a Chern-Simons topological field theory, and extension of the QHE to higher dimensions by Zhang and Hu \cite{Zhang:2001xs}, 
further developed in 
Refs.\cite{Bernevig:2003yz,Karabali:2002im,Karabali:2003bt,Karabali:2004km,Agarwal:2024fgp,Karabali:2026non,Hasebe:2014nia,Hasebe:2016tjg}, are specially relevant. The observation made in \cite{Zhang:2001xs,Zhang_2004} about the coincidences between algebra and geometry that make  monopoles and instantons, and these models of the QHE effect possible, namely the isomorphism between $S^1$ and $U(1)$ and between $S^3$ and $SU(2)$\footnote{The only spheres that are group manifolds are $S^0$, $S^1$ and $S^3$.}, will be important in what follows.

After the original derivation by Zhang et al of the effective theory in 2+1 dimensions from the microscopic theory, an alternative and very
elegant way to reach the same model and its consequences was found \cite{Frohlich:1991wb,Zee:1995avy}, very clearly presented by Zee in chapter VI.2 of his book on Quantum Field Theory \cite{Zee:2003mt}, based on general arguments of symmetry and dimensional 
analysis\footnote{In his book \cite{Zee:2003mt} Zee calls this approach "the short way" and the "what else can it be" method.}. 
The later approach yields most of the main results of the Fractional QHE in a very concise and beautiful way, 
that one would be hard-pressed to believe on, if not for the agreement with the experiment. 

Some lessons from this approach are (see Ch.VI.2 of \cite{Zee:2003mt}): 
i. the existence of composites that could be fermions or bosons, or fractional statistics (anyons), 
ii. the coefficient of the action $k$ has physical meaning as $k=\frac{1}{\nu}$, with $\nu $ the fraction of the Landau level occupied,
iii. the quasi-particles obey a fractional statistics with $\theta =\frac{\pi}{k}$, and it is an odd integer 
because a composite of $k$ anyons must be an electron (fermion), and 
iv. Chern-Simons field theory dynamics is key to describe  anyons.

I will review anyonic solutions to these effective theories in 2+1 dimensions, 
presented originally by Wilczek,Wu and Zee \cite{Wilczek:1981du,Wilczek:1982wy,Wilczek:1983cy,Wu:1984kd}, and then I will review the higher dimensional generalization of these theories and solutions due to Nepomechie, Wu and Zee \cite{Nepomechie1984,Wu:1988py}. 

The present work, originally motivated by String Theory and the formulation of background  independent and gauge invariant theories of extended objects, can also be thought of as a generalization of these models and solutions, including a dynamical space-time and non-abelian gauge symmetry (though in Nepomechie-Zee work an $SU(2)$ symmetry and quaternions were considered).

\subsection{ Case $d=2+1$}

This Subsection and the next follow \cite{Wilczek:1981du,Wilczek:1982wy,Wilczek:1983cy,Wu:1984kd}. Consider the action 
\begin{equation}
S=\int d^3x [g\epsilon ^{\mu\nu\rho}A_{\mu}F_{\nu\rho}+J_{\mu}A^{\mu}]+S_0,
\end{equation}
where $F_{\mu\nu}=\partial _{\mu} A_{\nu}-\partial  _{\nu}A_{\mu}$ and $S_0$ is the action of a massive free particle. Here $\eta _{\mu\nu}=diag(-,+,+)$ and $\epsilon ^{012}=+1$.
The field equation for $A_{\mu}$ is 
\begin{equation}
2g\epsilon ^{\mu\nu\rho}F_{\nu\rho}=-J^{\mu}.
\end{equation}
The current for a point-like source is 
\begin{equation}
J^{\mu}=e\int d\tau \delta ^{(3)}(x-y(\tau ))\frac{dy^{\mu}}{d\tau}(\tau).
\end{equation}
If the source is one particle at rest at the origin then $y^i(\tau )=0$, $y^0(\tau )=\tau $, 
then $J^i(x)=0$ and $J^0(x)=e \delta ^{(2)}(x,y)\equiv e \delta ^{(2)}(x^1,x^2) $, and then the equation of motion is
\begin{equation}
2g\epsilon ^{i_1i_2}F_{i_1i_2}=-e\delta ^{(2)}(\overrightarrow{x}),
\end{equation}
with $i_1,i_2=1,2$. Then 
\begin{equation}
A_i=-\frac{e}{8\pi g}\partial _i\phi
\end{equation}
where $\phi =\arctan{\frac{x^2}{x^1}}$ is the polar angle. 
As a differential 1-form $A=-\frac{e}{8\pi g}d\phi$. 

\subsection{Hopf $S^3\rightarrow S^2$ with spinors}

The action given above can be interpreted as an effective theory of solitons in a $O(3)$ nonlinear $\sigma $ model, 
where the  $\sigma $ field
is $n ^a$, with $a=1,2,3$ and $\sum _{a=1}^3n^a n^a =1$, and the conserved current  is
$J^{\mu}=\epsilon ^{\mu\nu\lambda}\epsilon _{abc}n^a\partial _{\nu}n^b\partial _{\lambda }n^c$.

Defining $n^a\equiv z^{\dagger}\sigma ^a z$,
with $a=1,2,3$ and with the spinor with two complex components $z_1, z_2$
\begin{equation}
z=
\left(\begin{array} {r}
z_1 \\  
z_2 
\end{array}\right)
\end{equation}
and
\begin{equation}
z^{\dagger}=
\left(\begin{array} {r r}
z^*_1 &,\; z^*_2 
\end{array}\right)
\end{equation}
with $z_1=x_1+ix_2$ and $z_2=x_3+i x_4$, and the condition $\mid z_1\mid ^2+\mid z_2\mid ^2=1$, that implies that $z$ 
parametrizes $S^3$. 
Then $n^a$ is real and one can verify that $\sum _{a=1}^3n^an^a=1$. 
That implies that the tip of $n^a$ parametrizes $S^2$. 
If we transform $z\rightarrow e^{i\theta}z$, $n^a$ remains invariant, 
therefore to any point in $S^2$ corresponds a circle in $S^3$.

\subsection{Generalization to higher dimensions}

In this subsection I will follow refs.\cite{Nepomechie1984,Wu:1988py}. The action proposed there is
\begin{equation}
S=\int d^{2m+1}x [g\epsilon ^{\mu _1...\mu _{2m+1}}A_{\mu _1...\mu _{m}}F_{\mu _{m+1}...\mu _{2m+1}}+J_{\mu _1...\mu _{m}}A^{\mu _1...\mu _{m}}]+S_0,
\end{equation}
where $F =dA$ and $S_0$ is the free action. Here $\eta _{\mu\nu}=diag(-,+,...,+)$ and $\epsilon ^{012...(2m+1)}=+1$. It is important that $m$ must be odd, otherwise the Chern-Simons term is a total derivative or an exact differential locally. That is 
$d(A^2)=(dA)A+(-1)^mA(dA)$, then if $m$ is even $d(A^2)=2AdA$, while if $m$ is odd $d(A^2)=0$. 
It follows that if $m$ is even $AF=AdA=d(A^2/2)$. 
If $m$ odd, then it is of the form $m=2k+1$. The extended objects have space-time dimension $m=2k+1$, that for $k=0$ corresponds to particles, for $k=1$ corresponds to branes, etc.. The space-time dimension is $d=2m+1=4k+3$, so we have particles in a $d=3$ space-time, membranes in a $d=7$ space-time, etc..
 
The field equation for $A_{\mu _1 \mu _m}$ is 
\begin{equation}
2g\epsilon ^{\mu _1...\mu _{2m+1}}F_{\mu _{m+1}...\mu _{2m+1}}=-J^{\mu _1...\mu _m}.
\end{equation}
The current for a singular brane-like source is 
\begin{equation}
J^{\mu _1\mu _m}=e\int d^m\sigma \delta ^{(2m+1)}(x-y(\sigma ^0,...,\sigma ^{m-1} ))\frac{\partial (y^{\mu _1},...,y^{\mu _{m}})}{\partial (\sigma ^0,...,\sigma ^{m-1} )}.
\end{equation}
If the singular brane is at rest with support in $x^j$ with $1\le j\le m-1$, $x^i=0$ if $m\le i\le 2m$, and we choose the gauge $A_{0j_1...j_{m-1}}=0$, 
that is possible because the gauge transformation $A'_m=A_m+d\Lambda _{m-1}$ leaves $F$ invariant, $F_{m+1}=dA_m=dA'_m$, and transforms the action by a total derivative, then the equations of motion are
\begin{equation}
2g\epsilon ^{j _1...j_{m-1}i_m...i_{2m+1}}F_{i_{m}...i _{2m}}=-e\epsilon ^{j _1...j _{m-1}}\delta ^{(m+1)}(x^i),
\end{equation}
with $1\le j_k\le m-1$ and $m\le i_k\le 2m$. Therefore
\begin{equation}
F_{i_{m}...i_{2m}}=-\frac{e}{2g}\frac{1}{(m+1)!} \epsilon _{i_m...i_{2m}}\delta ^{(m+1)}(x^i).
\end{equation}
Written as differential forms $F=\frac{F_{i_{m}...i_{2m}}}{(m+1)!}dx^{i_m}...dx^{i_{2m}}$, then
\begin{equation}
F=-\frac{e}{2g}\frac{1}{(m+1)!} \delta ^{(m+1)}(x^i)d^{m+1}x.
\end{equation}
The potential is
\begin{equation}
A=-\frac{e}{2g}\frac{1}{(m+1)!} \frac{d\Omega _m}{\omega _m}
\end{equation}
where $d\Omega _m$ is the volume element of the sphere $S^m$ of unit radius and 
\begin{equation}
\omega _m=\frac{2\pi ^{\frac{m+1}{2}}}{\Gamma(\frac{m+1}{2})}=\int _{S^m} d\Omega _m
\end{equation}
is its total volume.  
In the index notation the potential is
\begin{equation}
A_{i_{m}...i_{2m-1}}=-\frac{e}{2g}\frac{1}{(m+1)!\omega _m} \epsilon _{i_m...i_{2m}}\frac{x^{i_{2m}}}{r^{m+1}}.
\end{equation}
The exterior derivative of $A$, and hence $F$, is zero everywhere but in the world-volume of the brane, where it is singular, as it should. 
The integral of $F$ in any space-like volume containing the brane is, as it must,
\begin{equation}
 \int _{M^{m+1}}F= \int _{M^{m+1}}dA=\int _{\partial M^{m+1}}A=-\frac{e}{2g}\frac{1}{(m+1)!} \int _{S^m}\frac{d\Omega _m}{\omega _m}=-\frac{e}{2g}\frac{1}{(m+1)!}.
\end{equation}

\subsection{Quaternionic Formulation}

A construction similar to the one given in Section 3.3 is possible in seven dimensions with quaternions \cite{Wu:1988py} 
(see also \cite{Tze:1989jv}), where we have a 3-form topological current
$J^{\mu\nu\lambda}=\epsilon ^{\mu\nu\lambda\rho\sigma\tau\omega}\epsilon _{abcde}n^a \partial _{\rho}n^b\partial _{\sigma} n^c\partial _{\tau}n^d\partial _{\omega}n^e $, where $n^a$, with $a=1,...,5$ is the $\sigma$-field in a $SO(5)$ $\sigma$-model, and the Hopf term is $\int d^7x A_{\mu\nu\lambda}J^{\mu\nu\lambda}$. In this case we define two component quaternionic spinors
\begin{equation}
Q(x)=\left( \begin{array}{c} q_1(x) \\ q_2(x)\end{array} \right)
\end{equation}
with the condition $Q^{\dagger}Q=\overline{q}_1q_1+\overline{q}_2q_2=1$, and $n^a\equiv Q^{\dagger}\sigma ^aQ$, with the five $\sigma ^a$'s related to the standard Pauli matrices as
$\sigma _1\rightarrow \sigma _1$, $\sigma _2\rightarrow \sigma _2,\sigma _3, \sigma _4$ with the imaginary unit $i$ replaced by the quaternionic imaginary units $i, j, k$ respectively, 
and $\sigma _3\rightarrow \sigma _5$. It can then be verified that $n^an^a=1$. 
Notice that this is a explicit construction of the Hopf map $S^7\rightarrow S^4$, as the quaternionic spinor 
has eight real coordinates, and the condition $Q^{\dagger }Q=1$ defines $S^7$, while $n^an^a=1$ with $a=1,...,5$ defines $S^4$.

One can define the gauge potential $B_{\mu}=Q^{\dagger}\partial _{\mu}Q=\overline{q}_1\partial _{\mu}q_1+\overline{q}_2\partial _{\mu}q_2$, which is pure imaginary, as
$\overline{B}_{\mu}=-B_{\mu}$, and can therefore be identified as an $SU(2)$ potential. Under a right gauge transformation $q_1(x)\rightarrow q_1(x)q_0(x)$ and $q_2(x)\rightarrow q_2(x)q_0(x)$, with $q_0(x)$ a unitary quaternion $|q_0(x)|=1$, or $\overline{q}_0=q_0^{-1}$, $B_{\mu}$ transforms as a gauge field $B_{\mu}\rightarrow q_0^{-1}B_{\mu}q_0+q_0^{-1}\partial _{\mu}q_0$ . 
Notice that the $SU(2)$ generators are related to the imaginary quaternionic units as $i\rightarrow -i\sigma _1$, $j\rightarrow -i\sigma _2$ and $k\rightarrow -i\sigma _3$. 

In this formulation, the 3-form potential can be written as the Chern-Simons form of the potential, that is 
$A_{\mu\nu\lambda}=tr[B_{[\mu}G_{\nu\lambda ]}-\frac{2}{3}B_{[\mu}B_{\nu}B_{\lambda ]}]$, with the field potential 
$G_{\nu\lambda}=\partial _{\nu}B_{\lambda} -\partial _{\lambda} B _{\nu}+[B_{\nu},B_{\lambda}]$. 

In the notation of differential forms $A=\frac{1}{3!}A_{\mu\nu\lambda}dx^{\mu}dx^{\nu}dx^{\lambda}$ and $F=dA=tr[G^2]$. 
In quaternionic notation we only need to replace "tr" with "Re".

\section{Nepomechie-Wilczek-Wu-Zee models and Euclidean Chern-Simons branes}

\subsection{Generalization of the Wilczek-Wu-Zee model and Dirac Monopoles}

The gauge potential of a Dirac monopole is $A=-\frac{g}{4\pi }\cos
\theta \,  d \phi $, and the field strength is $F=dA=\frac{g}{4\pi
}\sin \theta\, d\theta\,  d\phi$ . This potential is ill-defined in
$\theta =0$ and in $\theta =\pi $, what can be fixed, as Wu-Yang did,
via gauge transformed potentials well defined in the northern and
southern hemispheres respectively, given by $A^N=\frac{g}{4\pi }[1- \cos \theta ]\,  d \phi  =A +d[\frac{g}{4\pi}\phi ]$ and
$A^S=-\frac{g}{4\pi }[1+ \cos \theta ]\,  d \phi  =A -d[\frac{g}{4\pi}\phi ]$. By Gauss magnetic law,
the integral of $F$ on a sphere is $\int _{S ^2}F =g$. In terms of the
potentials, Wu and Yang considered local charts
in both hemispheres, and obtained $\int _{S ^2}F =\int
_{S^1}[A^N-A^S]=\int _{S^1} 2\, d[\frac{g}{4\pi }\phi ]=g$, with $S^1$ the equator $\theta =\frac{\pi}{2}$.

Notice that the spin connection for the standard metric in $S^2$, vielbeins $e^1= d\theta$ and  $e^2=\sin \theta d\phi$, with zero torsion $T^a=de^a +\omega ^a_{\;b} e^b=0$ is $\omega^{12}=-\cos\theta d\phi $, 
proportional to the monopole potential, and that it transforms under $SO(2)\cong U(1)$ as the gauge potential under $U(1)$ gauge transformations. Also de curvature $R^{12}=\sin\theta d\theta d\phi $ is the area element in $S^2$.

Instead of thinking the QHE as charged particles on a plane with a magnetic field perpendicular to the plane, 
Haldane \cite{Haldane:1983xm} proposed to think the QHE as charged particles on a very large sphere with 
a magnetic field generated by a magnetic monopole at the center of the sphere. In the spirit of Haldane,  
we could consider the limit
of a very large sphere, or a very small neighbourhood of the north pole, with $\theta \ll 1$, 
to recover the Euclidean
Wilczek-Wu-Zee model \cite{Wilczek:1981du,Wilczek:1982wy,Wilczek:1983cy,Wu:1984kd}, 
with the radial coordinate as the euclidean time.

It is interesting that then the anyon/flux tube may be thought of as a Dirac string, ending at the monopole 
in one end\footnote{That in the limit of a sphere of very large radius goes to minus infinity euclidean time.}, and at infinity in the other.  The action of \cite{Wilczek:1981du,Wilczek:1982wy,Wilczek:1983cy,Wu:1984kd} can, in the singular case, be seen as an action for Dirac strings in 2+1 dimensions, somewhat analogous to Dirac \cite{Dirac:1931kp,Dirac:1948um} and Hull \cite{Hull:2024uwz,Hull:2025yww} works. 

If $\theta\ll 1$ then $\cos \theta \approx 1-\frac{\theta^2}{2}$, and to first orden in $\theta $ we have $A^N =0$ and
$A^S=-\frac{g}{2\pi }\,  d \phi $, which is singular at the Dirac string at $\theta =0$. If we calculate the same integral as above
 but for $S^1$ a small circle with
constant $\theta\ll 1$, then again $\int _{S ^2}F =\int _{S^1}[-A^S]=\int _{S^1} 2\, d[\frac{g}{4\pi }\phi ]=g$. 
Notice that to go to the Wilczek-Wu-Zee model we must change the notation and replace $g\rightarrow
\frac{e}{4g}$.

\subsection{Membranes in seven dimensions}

If the branes have no boundaries, the Transgression action of \cite{Mora:2023ngp} 
in seven dimensions and for gauge group $SO(8)$ reduces to
$$S=c_1\int _{M^7} F_4 A_3+c_2\int _{M^7}  \, ^*j A_3,$$
where $F_4\equiv STr[F^2]$ and $A_3\equiv \mathcal{C}_{3}(A,F)$, so that $F_4= d A_3$, and $\, ^*j$ is a 4-form 
that has as its Hodge dual  the 3-form current $j$. 
Then $ \int _{M^7}  \, ^*j A_3=\int _{M^7}d^7x \: j^{\mu _1\mu _2 \mu _3}A_{\mu _1\mu _2 \mu _3}$
In the distributional case
$$j ^{\mu _1\mu _2\mu _3} (x)=q \int _{M_3}d^3\xi \epsilon ^{a_1 a_2 a_3} \frac{\partial X^{\mu _1}}{\partial \xi ^{a_1}} \frac{\partial X^{\mu _2}}{\partial \xi ^{a_2}} \frac{\partial X^{\mu _3}}{\partial \xi ^{a_3}}\delta (x-X(\xi ))$$
we have
$$ \int _{M^7}d^7x \: j^{\mu _1\mu _2 \mu _3}A_{\mu _1\mu _2 \mu _3}=q \int _{M_3}d^3\xi \epsilon ^{a_1 a_2 a_3} \frac{\partial X^{\mu _1}}{\partial \xi ^{a_1}} \frac{\partial X^{\mu _2}}{\partial \xi ^{a_2}} \frac{\partial X^{\mu _3}}{\partial \xi ^{a_3}}A_{3\;\mu_1\mu_2\mu _3}(X(\xi ))=q \int _{M^3}A_3,$$
or
$$\int _{M^7}  \, ^*j A_3=q \int _{M^3}A_3.$$
This expressions are similar to euclidean version of the Nepomechie-Wu-Zee model \cite{Nepomechie1984,Wu:1988py} 
in its formulation with $SU(2)$ (or equivalently quaternionic) gauge fields, but it has a larger $SO(8)$ gauge symmetry,
both for the background and the branes. 

Note that it was not possible in the previous Subsection to extend the euclidean
Wilczek-Wu-Zee three-dimensional action \cite{Wilczek:1981du,Wilczek:1982wy,Wilczek:1983cy,Wu:1984kd} to a gauge model with $SO(4)$ gauge symmetry, as the worldline action for the anyons should have 
been proportional to $Tr[T^A]$, with $T^A$ the generators of $SO(4)$, which is zero. Also the factorization property of the trace does not hold in this case.

On finding the equations of  motion for our seven dimensional action, we impose the condition $\delta S =0$, but we will just consider variations under  $\delta A_3$, not under $\delta A$, because the former factorizes, and its coefficient equated to zero
will be the equations of motion sought. We have $\delta [F_4 A_3]=[\delta F_4 ] A_3+F_4\delta A_3=d[\delta A_3]A_3+F_4 \delta A_3$. 
But $d[\delta A_3]A_3=+[\delta A_3]d A_3+d[(\delta A_3) A_3]=+[\delta A_3] F_4 +d[(\delta A_3) A_3]$. 
Then, ignoring total derivatives because we are dealing with branes without boundaries, we get $\delta [F_4 A_3]=2 F_4 \delta A_3$.
Assuming the 4-form $\, ^* j$ is defined independently of $A$ (that is, independently of the vielbein, or the metric), which will not be true for its Hodge dual $j$, 
we have $\delta [\, ^* j A_3]=\, ^* j\delta A_3$. Then the variation of the action (without boundary terms) is

$$\delta S=\int _{M^7}[2 c_1 F_4+c_2 \, ^* j]\delta A_3 =0$$

and the equation of motion is

$$F_4=-\frac{c_2}{2 c_1}\, ^* j$$

If we write the variations in terms of $A$ and $F$, we have  $F_4=STr[F^2]$ and $A_3=STr[AdA+\frac{2}{3}A^3]$, and  $\delta A_3=2 STr[F\delta A]$, we see that $F=0$ is a sort of degenerate limit, as
in that case $\delta A_3=0$, but we will consider it a particular case of the generic $F\neq 0$ case. 

If the branes had boundaries, one would have non-trivial conditions on those boundaries, even in the $F=0$ case.

\subsection{Solution for Euclidean Chern-Simons branes analogous to the Nepomechie-Wu-Zee static solution}

A particular solution for a background and a branes corresponds to the seven dimensional sphere $S^7$,  which has $SO(8)$ gauge curvature 
$F=0$,
 and therefore $F_4=0$, with a Dirac delta singularity associated to the Dirac string, in this case the Dirac brane, 
where the gauge potential is ill-defined (for instance the north pole in a chart well defined at the south pole). 
Again, as in the case of the monopole, the limit of large radius of the sphere, or coordinates close to the north pole, corresponds to the solution of \cite{Nepomechie1984,Wu:1988py}.
Notice that the Hopf fibration non triviality is essential , as $S^7\neq S^4\times S^3$, and only the former manifold satisfies $F=0$. 

The Dirac-like brane described is not pure gauge because of the topological obstruction expressed in the non vanishing of the Pontryagin numbers of the self-dual and anti self-dual parts of the spin connection, calculated explicitly in the next subsection in a way somewhat different from the calculation in the original BPST paper \cite{Belavin:1975fg}, and in the standard literature 
(see for instance \cite{Zee:2003mt})\footnote{Also because the Hopf fibration is topologically non trivial. Notice that in the explicit expression of the Hopf map at the end of Subsection 3.5 we see that the $SU(2)$ invariance resulting from the right multiplication for the unitary quaternion $q_0(x)$ does not affect the $S^4$ coordinates $n^a$, meaning that one have the fibre $SU(2)\cong S^3$ 
at each point $x$ of $S^4$. That means that one of the two $SU(2)$'s of $SO(4)\cong SU(2)\times SU(2)$ is to be thought as the $S^3$ fiber over $S^4$.}. Here  I use a variation of that method, using a small radius three-sphere $S^3$ surrounding the north pole (the brane), instead of the equator of $S^4$, as it is usually done. This is analogous to the computation described above for the magnetic monopole. 
The conclusion, already clear from the equation of motion $F_4=-\frac{c_2}{2 c_1}\, ^* j$, is that $F_4$ is zero everywhere except on the brane, where it is singular.

\subsection{Spin connection and curvature of $S^4$, self-dual and anti-self-dual components and the BPST instanton}

The spin connection of $S^4$ can be thought of as a $SO(4)$ gauge potential, and the Riemann curvature as its curvature or field strength. 
As $SO(4)\cong SU(2)\times SU(2)$ , it can be decomposed in a self-dual and an anti-self-dual part as
\begin{equation}
^{\pm}\omega ^{ab}=\frac{1}{2}[\omega ^{ab}\pm\frac{1}{2}\epsilon^{ab}_{\;\; cd}\omega ^{cd}]
\end{equation}
and 
\begin{equation}
^{\pm} R^{ab}=\frac{1}{2}[R ^{ab}\pm\frac{1}{2}\epsilon^{ab}_{\;\; cd} R^{cd}]
\end{equation}
Then $\omega ^{ab}=^+\omega ^{ab}+ ^-\omega ^{ab}$ and $R^{ab}= ^+R^{ab}+ ^- R^{ab}$. The last equality follows from the fact that the generators of one of the two $SU(2)$'s,
conmute with the generators of the other one. Schematically, if $A= ^+A + ^-A$ then
$$F= dA+ A^2= d ^+A + d^-A +^+A^2+ ^-A^2 + ^+A ^-A +^-A ^+A = ^+F+ ^-F,$$
remembering $ ^{\pm}A$ are 1-forms, and $ ^+A ^-A +^-A ^+A =[^+A ,^-A ] = 0$.
Considering that $R^{ab}=e^a e^b$, then
\begin{equation}
^{\pm} R^{ab}=\frac{1}{2}[e^{a} e^{b}\pm\frac{1}{2}\epsilon^{ab}_{\;\; cd} e^{c} e^{d}].
\end{equation}

The self-dual and anti self-dual parts of the spin connection of $S^4$ are equal to the $SU(2)$ gauge potential of an instanton and an anti
instanton respectively, just as the spin connection of $S^2$ is the $U(1)$ gauge potential of a monopole. What is presented in this subsection is the "spherical version" of the $SU(2)$ instanton, while in the appendix the "stereographic version" is reviewed. This is analogous to the spherical version of the monopole described for instance in Ch.IV.4 of \cite{Zee:2003mt}, versus the stereographic version of the appendix.

The Chern Character of these curvatures can be computed from
$$ch(F)=STr\left[ \exp{\left( i \frac{F}{2\pi }\right) }\right]\equiv \sum _{k=0}^{\infty}ch _k(F)$$
where
$$ch _k(F)=\frac{i^k}{(2\pi )^k k!}  STr \left[  F^k\right] .$$
In particular the first Pontryagin class is minus the second Chern character
$$p_1(F)=-c_2(F)=\frac{1}{8\pi ^2} STr[F^2]$$
If $F=\frac{1}{2}R^{ab}J_{ab}$ and $STr[J_{ab}J^{cd}]=\frac{1}{2}[\delta _a^c \delta _b^d-\delta _a^d \delta _b^c]$, then $STr[F^2]=\frac{1}{4}R^{ab}R_{ab}$.
We have, considering that $e^a e _a=0$, 
$$ ^{\pm}R_{ab}   ^{\pm}R ^{ab}=\pm \frac{1}{2}\epsilon _{abcd} e^a e^b e^c e^d.$$
The Pontryagin number is a topological invariant given by
$$q=\int _{S^4}p_1(F)=\pm \frac{1}{8\pi ^2} \frac{1}{4}  \frac{1}{2} \int _{S^4} \epsilon _{abcd} e^a e^b e^c e^d.$$
Explicitly $\int _{S^4} \epsilon _{abcd} e^a e^b e^c e^d=4! \int _{S^4} e^1 e^2 e^3 e^4=4! \frac{8}{3}\pi ^2$, because the volume of the unit radius $S^4$ 
with the volume element induced by its embedding in $R^5$ is $\frac{8}{3}\pi ^2$. Therefore the Pontryagin number, also known as the instanton number, is
$$ q=\pm 1.$$
It is worthwhile to notice that for the whole Riemann curvature $R^{ab}R_{ab}=e^a e^b e_a e_b=0$, and in that case $q=0$ . 
Also note that the Euler number in that case is the one of the four-sphere
$$\chi (S^4)=\frac{1}{2} \frac{1}{(4\pi )^2}\int _{S^4}\epsilon _{abcd}R^{ab}R^{cd}= \frac{1}{2} \frac{1}{(4\pi )^2}\int _{S^4}\epsilon _{abcd} e^{a} e^{b} e^{c} e^{d}=\frac{1}{2} \frac{1}{(4\pi )^2}4!\frac{8\pi ^2}{3}=2.$$

\subsection{Exotic Spheres}

Exotic spheres, that are homeomorphic but not diffeomorphic to standard spheres,  were discovered by Milnor \cite{Milnor} in 1956 in seven dimensions, and afterwards in other dimensions. Their possible relevance to Physics in diverse contexts has been studied in several works, 
old and new 
\cite{Yamagishi:1983dw,Freund:1984am,Witten:1985xe,Brans:1992mj,Brans:1994hq,Brans:1994wv,Brans:1996ix,Asselmeyer:1996bh,Asselmeyer-Maluga:2015qka,Gherardini:2023uyx,Berman:2024yoa,SchettiniGherardini:2025ydb}. 
One interesting observation is the "Brans conjecture" \cite{Brans:1994hq,Brans:1994wv,Brans:1996ix,Asselmeyer:1996bh}, stating that, 
in the context of the theories considered in those works, the Exotic differential structure would manifest as 
sources of the gravitational field (particles and extended objects), as a sort of "twist of emptiness". It is amusing that in the theoretical context of the present paper, as shown in the previous subsections, we may interpret particles and branes as a consequence of the non trivial topology of
standard "garden variety" spheres. The question naturally arises of whether Exotic spheres in seven 
dimensions would correspond to solutions to our theory, as standard spheres do. Here a difficulty is that for 
Exotic spheres $R^{ab}\neq \frac{1}{l^2}e^a e^b$ \cite{Berman:2024yoa}, 
then $F\neq 0$, therefore, while they can still be solutions, it would take some more work to prove it.

It is encouraging that multi-instanton solutions may be identified with Exotic spheres as Yamagishi \cite{Yamagishi:1983dw} states: 
"The seven-dimensional sphere constructed by Milnor is the total space of the $S^3$-bundle over $S^4 $ characterized 
by the mapping $f_{kj}: g\mapsto \left(\frac{x}{|x|}\right) ^k g \left(\frac{x}{|x|}\right)^j$ $[f_{kj}\in SO(4)]$.    Here $k$ and $j$ are integers satisfying
$k+j=1$, wich is needed for the bundle to coincide with $S^7$ topologically. Remarkable fact is that these spheres with non-zero value of 
$\lambda \equiv [(k-j)^2-1]/8\, (mod \, 28)$  have distinct differential structures from the ordinary $S^7$ . They are denoted $\Sigma ^7$.  From the preceding argument
 of $SU(2)$ instantons, we can recognize that $k$ and $-j$ are the Pontryagin indices corresponding to those of two $SU(2)$ contents of $SO(4)$ instantons.
 Therefore, if we construct $\Sigma ^7$ by $SO(4)$ instantons, the total instanton number $(k-j)$ reflects the exoticness $\lambda$ of the sphere.
 A particular case$(k,j)=(1,0)$ or $(0,1)$ corresponds to the Hopf-fibering of $S^7$,  which is diffeomorphic to the ordinary sphere, i.e. $\lambda =0$."
 
One may heuristically think of multiple branes located very far away of each other and with different charges as multi-instanton solutions, with suitable identifications at the equator to make the configuration regular, as a possible solution 
that may correspond to Exotic spheres.
The fact that the equations in seven dimensions and for  branes are non linear, unlike the case of three dimensions and 
anyons, makes the problem harder. Still, multi-instanton solutions are known in exact 
form \cite{Atiyah:1978ri,atiyah1979geometry}.

If exotic spheres in seven dimensions do correspond to solutions of our theory remains an interesting open problem.
 
\section{Lorentzian Chern-Simons brane solution}

In the Lorentzian case again it is enough (though not necessary) for a configuration to be a solution for our brane action that $F=0$ and that it would be singular in the location of the brane. We are interested in a 2-branes in a seven dimensional space. Explicit static solutions have been known for some time, introduced by Ba\~nados in ref.\cite{Banados:1997df} interpreted as constant curvature black holes,  
and then reinterpreted in some cases as brane solutions and further investigated by 
Guilleminot, Olea and Petrov in ref.\cite{Guilleminot:2017hfy}. In the next subsection I will review these works, and in the following one I will check that it has the requiered singularity at the brane.

\subsection{Review of Constant Curvature Black Holes (CCBH) and Brane Solutions}

In this review Subsection I follow Section 3 of Ref.\cite{Guilleminot:2017hfy}. I will use the convention $l=1$ for the arbitrary scale parameter in this subsection and the next. To recover the parameter $l$ one must 
replace $e\rightarrow e/l$ for the vielbein and $r\rightarrow r/l$ for the radial coordinate.

Consider the line element
\begin{equation}
ds^2=A^2(r)\eta _{ij}\hat{e}^{i}_{\mu}\hat{e}^{j}_{\nu}dx^{\mu}dx^{\nu}+B^2(r)dr^2+r^2\delta _{mn}\tilde{e}^m_{\alpha}\tilde{e}^n_{\beta}dx^{\alpha}dx^{\beta}
\end{equation}
where $r$ is the radial coordinate, the brane metric is $\hat{g}_{\mu\nu}=\eta _{ij}\hat{e}^{i}_{\mu}\hat{e}^{j}_{\nu}dx^{\mu}dx^{\nu}$ with $\mu ,\nu , i,j=0,...,p$ and the transverse 
metric $\tilde{g}_{\alpha\beta}=\delta _{mn}\tilde{e}^m_{\alpha}\tilde{e}^n_{\beta}$ with $\alpha ,\beta ,m,n=p+1,...,d$, and $A(r)$ and $B(r)$ are functions to be determined.
The corresponding vielbeins are
\begin{equation}
e^i=A(r)\hat{e}^i=A(r)\hat{e}^i_{\mu}dx^{\mu}  ,  e^r=B(r)dr ,  e^n=r\tilde{e}^n=r\tilde{e}^n_{\alpha}dx^{\alpha}.
\end{equation}
The spin connection $\omega ^{ab}=-\omega ^{ba}$ is determined if we impose that the torsion be zero, that is $T^a=de^a+\omega ^{ab}e_b=0$, in which case
\begin{equation}
\omega ^{mi}=0  ,  \omega ^{ij}=\hat{\omega }^{ij}  ,  \omega ^{mn}=\tilde{\omega }^{mn}  ,  \omega ^{ir}=\frac{A'(r)}{A(r)B(r)}e^i  ,  \omega ^{mr}=\frac{e^m}{rB(r)} .
\end{equation}
The Riemann curvature $R^{ab}=d\omega ^{ab}+\omega ^{ac} \omega _c^{ \; b}$ is
\begin{eqnarray}
R^{ij}=\hat{R}^{ij}-\frac{A'^2}{(AB)^2}e^ie^j \\
R^{im}=-\frac{A'}{AB^2}e^ie^m \\
R^{ir}=-\frac{1}{AB}\left( \frac{A'}{B}\right)'e^ie^r \\
R^{rm}=\frac{B'}{rB^3}e^re^m \\
R^{mn}=\tilde{R}^{mn}-\frac{1}{(rB)^2}e^me^n
\end{eqnarray}
where the $\hat{R}^{ij}=d\hat{\omega } ^{ij}+\hat{\omega } ^{ik} \hat{\omega }_k^{ \; j}$ is the intrinsic Riemann curvature of the brane, 
and $\tilde{R}^{mn}=d\tilde{\omega } ^{mn}+\tilde{\omega }^{ml} \tilde{\omega }_l^{ \; n}$ the intrinsic Riemann curvature of the transverse manifold.

The constant curvature solutions correspond to $F=0$, with the gauge field strength $F$ given by 
\begin{equation}
F=\frac{1}{2}\left[R^{ab}+e^ae^b\right]J_{ab}+T^aP_a.
\end{equation}
By construction the torsion of the configurations considered is zero $T^a=0$. We must then impose $R^{ab}+e^a e^b=0$. From the previous expressions it results
\begin{equation}
A(r)=\frac{1}{B(r)}=\sqrt{r^2+C},
\end{equation}
where $C$ is an integration constant, and it must be that
\begin{eqnarray}
\hat{R}^{ij}+C \hat{e}^i\hat{e}^j=0\\
\tilde{R}^{mn}-C\tilde{e}^m\tilde{e}^n=0
\end{eqnarray}
which implies that both the brane and the transverse manifold have constant curvature.

If $C>0$ the brane is a Lorentzian space of constant negative curvature. 

If $C=0$ we get AdS. 

If $C<0$ the brane would be a Lorentzian space with positive curvature, that is a dS space, and the transverse space would be 
a Euclidean space with negative curvature. That is impossible, except for a $(d-2)$-brane with world-volume dimension $d-1$ in a space-time of dimension $d+1$, for in that case the transverse space is one dimensional and has no intrinsic 2-form curvature.

\subsection{Transgression brane solution, angular defect and conic singularity}

As said above, particular solutions in our Chern-Simons branes models consist of 
configurations that satisfy both the condition $F=0$ and that are singular at the location of the branes, 
as seen above in the Euclidean case. The CCBH solution \cite{Banados:1997df,Guilleminot:2017hfy} satisfies the condition $F=0$, as shown in the previous Subsection.
In this Subsection I will verify that in the case $C>0$ the CCBH has the required singularity at the location of the brane.
The case of interest for us are 2-branes (three dimensional world volume) in a seven dimensional space.

As an easier example, consider the two dimensional conic metric 
\begin{equation}
ds^2=\frac{dr^2}{(1-\alpha )^2}+r^2d\phi ^2.
\end{equation}
With the variable change $r'=\frac{r}{(1-\alpha )}$ it becomes
\begin{equation}
ds^2= dr'^2 +(1-\alpha )^2r'^2d\phi ^2= dr'^2 +r'^2d\phi '^2,
\end{equation}
where $d\phi '=(1-\alpha)d\phi$. The last expression is similar to the flat metric, but when $\Delta \phi =2\pi$ there is an angular defect $\Delta \phi '=(1-\alpha )2\pi$ corresponding to 
the conic singularity. If $C=1-\alpha $, the vielbein of the first metric is $e^1=\frac{dr}{C}$ and $e^2=r d\phi$, and the zero torsion condition gives the spin connection $\omega ^{12}=-C d\phi $. 
The curvature is $R^{12}=d \omega ^{12}=0$, but intuitively we expect a curvature singularity at $r=0$.  To verify that, we consider the Euler Characteristic for a two-dimensional manifold with boundary (see for instance \cite{Eguchi:1980jx})
$$\chi (M^2)=\frac{1}{4\pi }\int _{M^2}\epsilon R -\frac{1}{4\pi }\int _{\partial M^2}\epsilon\omega +\frac{1}{2\pi }\sum _i(\pi -\theta _i),$$
where $\theta _i$ are the angles of the tangents at the vertices of the boundary, if there is any. Eguchi et al. compute the invariant for a disc, in three versions, namely a flat disc, a hemisphere and a flat square, and the result is $\chi (disc)=1$. If we do the same calculation with the conical metric the result is $\chi =C=1-\alpha $, coming from the second term, while the first and third term are zero. Yet, as the result for this topological invariant must be $1$, we must have $\frac{1}{4\pi }\int _{M^2}\epsilon R =\alpha $, that means that 
$R^{12}=2\pi \alpha \delta (\mathbf{r})dx dy $, as expected.

Similarly, with the metric
\begin{equation}
ds^2=\left[C+\frac{r^2}{l^2}\right]\eta _{ij}\hat{e}^{i}_{\mu}\hat{e}^{j}_{\nu}dx^{\mu}dx^{\nu}+\frac{dr^2}{\left[C+\frac{r^2}{l^2}\right]}+r^2\delta _{mn}\tilde{e}^m_{\alpha}\tilde{e}^n_{\beta}dx^{\alpha}dx^{\beta}
\end{equation}
if we choose $r'=\frac{r}{\sqrt{C}}$, then for $r$ and $r'$ small the metric is
\begin{equation}
ds^2\approx C\eta _{ij}\hat{e}^{i}_{\mu}\hat{e}^{j}_{\nu}dx^{\mu}dx^{\nu}+dr'^2+C r'^2\delta _{mn}\tilde{e}^m_{\alpha}\tilde{e}^n_{\beta}dx^{\alpha}dx^{\beta},
\end{equation}
that correspond to an angular defect of $\sqrt{C}$ and the corresponding conical singularity at $r=r'=0$.

For a more explicit calculation, I will consider these metric at the lowest non vanishing order for $r\rightarrow 0$, that is
\begin{equation}
ds^2=\zeta ^2 \eta _{ij}\hat{e}^{i}_{\mu}\hat{e}^{j}_{\nu}dx^{\mu}dx^{\nu}+\frac{dr^2}{\zeta ^2}+r^2\delta _{mn}\tilde{e}^m_{\alpha}\tilde{e}^n_{\beta}dx^{\alpha}dx^{\beta},
\end{equation}
where $C=\zeta ^2$, and the metric transverse to the brane in this limit, corresponding to $d x^{\mu }=0$, is 
\begin{equation}
ds^2=\frac{dr^2}{\zeta ^2}+r^2\delta _{mn}\tilde{e}^m_{\alpha}\tilde{e}^n_{\beta}dx^{\alpha}dx^{\beta}.
\end{equation}
From this metric the vielbein is $e^r=\frac{d r}{\zeta}$ and $e^m=r \tilde{e}^m$, and the zero torsion spin connection is $\omega ^{r m}=-\zeta \tilde{e}^m$ and 
$\omega ^{mn}=\tilde{\omega }^{mn}$. The components of the curvature are $R^{rm}=-\zeta T^m=0$ and $R^{mn}=\tilde{R}^{mn}-\zeta ^2\tilde{e}^m \tilde{e}^n$. 
For us this space is four dimensional, with the coordinates $r$ and $x^{\alpha}$ with $\alpha =1,2,3$ the coordinates of the three sphere. To verify the existence of a delta function singularity in the origin we can, as in the two dimensional case, consider the Euler Characteristic of on open ball, which can be proved to be $\chi (B^4)=1$. The Euler Characteristic of a four dimensional manifold with boundary (assumed smooth) is \cite{Eguchi:1980jx}
$$\chi (M^4)=\frac{1}{2}\frac{1}{(4\pi )^2} \left[ \int _{M^4} \epsilon R^2-\int _{\partial M^4}\epsilon\left( \omega d\omega +\frac{2}{3}\omega ^2\omega\right) \right] $$
For the metric considered (without considering the singularity at the origin) $ \int _{M^4} \epsilon R^2=0$, and from the contribution of the boundary term we get
$$\chi =\frac{\zeta }{2}\left[3-\zeta ^2\right].$$
This expression gives the correct value $\chi =1$ in the case without a conical singularity, $\zeta =1$. If there is a conical singularity the only way to have $\chi =1$ is to have 
a delta function singularity of the curvature at $r=0$, such that
$$ \frac{1}{2}\frac{1}{(4\pi )^2}  \int _{M^4} \epsilon R^2= 1-\frac{\zeta }{2}\left[ 3-\zeta ^2\right] ,$$
which is zero, as it should, if $\zeta =1$, and implies that 
$$\epsilon R^2 =2(4 \pi )^2[1-\frac{\zeta }{2} (3-\zeta ^2)]\delta _{M^3}$$
where $\delta _{M^3}$ is a Dirac delta differential form such that when integrated on the coordinates normal to the 3-brane at the origin gives 
$$\int _{M^4}\delta _{M^3}=1$$
if the origin is in the integration domain.

\section{Quantum phases and statistics}

The unity of Nature may seem self-evident to anyone who has though about these questions, 
as any particle or any empty little cube of space somehow "knows" how to act in any circumstance, and so it seems to "know" the totality of physical law. On the matter of quantum statistics, or the division of particles in fermions (matter) and bosons (interaction), there are three ways to unify those statistics, namely supersymmetry (or the existence of a symmetry of the classical action that mixes both types), 
that at a fundamental level there are only fermions (and bosons are composite), or
that a fundamental level there are only bosons (and fermions are composite, with quantum phases coming from topology).

Here we consider the third option. Once again, the Quantum Hall Effect in 2+1 dimensions is a source of insight, as an example of composite fermions and dualities between bosons and fermions, and dualities between vortices and fundamental excitations (see \cite{Zee:2003mt} for a nice review of all these aspects). 
This third line of thought has already several decades, with important works being Refs.\cite{Jackiw:1975fn,Jackiw:1976xx,Hasenfratz:1976gr} and Refs.\cite{Zee:1989fr,Zee:1992va}.

\subsection{Review of Nepomechie-Wilczek-Wu-Zee models and quantum phases and statistics}

The interchange of two particles in 2+1 dimensions gives a phase factor of $\exp (\frac{- i e}{4 g})$ that corresponds to fractional statistics as it was shown by Wilczek, Wu and Zee \cite{Wilczek:1981du,Wilczek:1982wy,Wilczek:1983cy,Wu:1984kd}.  
This was shown first for two static vortex solutions and then with more general arguments.

The starting point is the so called central identity of QFT \cite{Zee:2003mt}: $\int \mathcal{D}\phi  \exp [-\frac{1}{2}\phi K \phi +J\phi]=\exp [\frac{1}{2}JK^{-1}J]$.

In $d=2+1$ we have $K^{\mu\rho }=2g\epsilon ^{\mu\nu\rho}\partial _{\nu }$, and $K^{-1}_{\mu\rho}=\frac{1}{2 g}\epsilon _{\mu\nu\rho} \frac{\partial ^{\nu}}{\partial ^2}$. 
We must use the gauge invariance $A_{\mu}\rightarrow A_{\mu}+\partial _{\mu}\xi $ to impose the Lorentz gauge condition $\partial _{\mu}A^{\mu}=0$ for $K$ to be invertible and $K^{-1}$ to exist.
Then 
$$\int \mathcal{D}A \exp[i\int d^3 x(g\epsilon ^{\mu\nu\rho } A_{\mu}F_{\nu\rho}+J^{\mu}A_{\mu})]=\exp \left[ \frac{i}{4 g}\int d^3 x\int d^3 yJ^{\mu}(x)\epsilon _{\mu\nu\rho }\frac{\partial ^{\nu}}{\partial ^2}J^{\rho}(y)\right] .$$
There $ \int d^3 x\int d^3 yJ^{\mu}(x)\epsilon _{\mu\nu\rho }\frac{\partial ^{\nu}}{\partial ^2}J^{\rho}(y)$ is the so called  "Hopf term". Notice that in $2+1$ we have that $\frac{\partial _{\nu}}{\partial ^2}\sim \frac{x_{\nu}}{\mid x\mid ^3}$.
For a distributional current $J^{\mu}(x)=e\int _{\mathcal{C}} d\tau \delta ^{(3)}(x-y(\tau ))\frac{dy^{\mu}}{d\tau }$ the Hopf term is 
$e\int _{\mathcal{C}_1}d x^{\mu}\int _{\mathcal{C}_2}dy^{\nu}\epsilon _{\mu\nu\rho} \frac{(x-y)^{\rho}}{\mid x-y \mid ^3}=e 4\pi  \mathit{lk}(\mathcal{C}_1,\mathcal{C}_2)$. 
There $\mathit{lk}(\mathcal{C}_1,\mathcal{C}_2)$ is the Gauss linking number, a topological invariant, and the quantum phase associated to these trajectories is
$ \exp [\frac{ie\pi }{g} \mathit{lk}(\mathcal{C}_1,\mathcal{C}_2)]$.

In $d=6+1$ and for the model of Nepomechie-Wu-Zee \cite{Nepomechie1984,Wu:1988py} we have 
$$K^{\mu _1\mu _2\mu _3,\mu _5\mu _6\mu _7}=2 g \epsilon ^{\mu _1\mu _2\mu _3\mu _4\mu _5\mu _6\mu _7}\partial _{\mu _4}=
-2 g \epsilon ^{\mu _1\mu _2\mu _3\mu _5\mu _6\mu _7\alpha }\partial _{\alpha }.$$
Schematically, $K=2 g \epsilon \partial $, therefore $K^{-1}=\frac{1}{2g}\epsilon \frac{\partial }{\partial ^2}$, where had to use the gauge invariance $A_3\rightarrow A_3+d\xi _2$ to impose the Lorentz gauge condition $\partial _{\mu }A^{\mu\nu\rho}=0$ for $K$ to be invertible. Explicitly 
$$(K^{-1})_{\nu _1\nu _2\nu _3,\nu _5\nu _6\nu _7}=-\frac{1}{2 g}\frac{1}{4!}\epsilon _{\nu _1\nu _2\nu _3 \nu _5\nu _6\nu _7\beta }\frac{\partial ^{\beta}}{\partial ^2}.$$

As in $2+1$ case, if we integrate out $A_3$ we obtain an action proportional to the Hopf term, that in this case is
$$\int d^7x\int d^7y J_{\mu\nu\lambda}(x)\epsilon ^{\mu\nu\lambda\rho\sigma\tau\omega }\frac{\partial _{\rho}}{\partial ^2}J_{\sigma\tau\omega }(y).$$
For singular branes 
$$J^{\mu\nu\rho }(x)=\int d^3\xi \; \delta ^{(7)}(x-y(\xi ^0,\xi ^1, \xi ^2))\frac{\partial (y^{\mu},y^{\nu}, y^{\rho})}{\partial(\xi ^0,\xi ^1,\xi ^2)}$$
and considering that in $6+1$ we have that $\frac{\partial _{\nu}}{\partial ^2}\sim \frac{x_{\nu}}{\mid x\mid ^7}$,
the Hopf invariant is a generalized linking number between 2-branes in seven dimensional space
$$\oint \oint dy^{\mu\nu\lambda} dz^{\sigma\tau\omega }\epsilon ^{\mu\nu\lambda\rho\sigma\tau\omega}\frac{(y-z)_{\rho}}{\mid y-z\mid ^7}$$
where  $dy^{\mu\nu\lambda} =d\xi ^0 d\xi ^1 d\xi ^2 \frac{\partial (y^{\mu},y^{\nu}, y^{\lambda })}{\partial(\xi ^0,\xi ^1,\xi ^2)}$. 

The topological quantum phase resulting from interchanging two 2-branes is proportional to the Hopf invariant, which is a topological invariant dependent on the homotopy class of the world volume of both branes, and can be interpreted as a higher dimensional generalization of quantum statistics, from point particles in 2+1 to 2-branes in 6+1. The issue of Quantum Phases and Quantum Statistics in 
this kind of model was also investigated in depth by Tze and Nam \cite{Tze:1989jv}.

\subsection{Quantum phases for Chern-Simons branes}

 The quantum phase associated to the interchange of two closed Chern-Simons 2-branes with the initial and final time identified is just the action (or $S/\hbar $ if $\hbar$ is not chosen as 1), given by
 $S=c \int _{\mathcal{M}^3}\mathcal{C}_{3} $ (where $c$ is a constant), 
which is a topological invariant for closed manifolds (that is, if the branes have no boundaries, as we assume), 
and it is non trivial\footnote{The three-dimensional Chern-Simons form satisfy $STr[F^2]=d \mathcal{C}_{3}$, and in general $STr[F^n]$ is associated with the Homotopy group $\pi_{2n-1}(G)$, where $G$ is the gauge group used to define the curvature $F$ (see Appendix A of \cite{Zumino:1983rz} ). 
For us $n=2$ and $G=SO(8)$ in the Euclidean case and $G=SO(6,2)$ in the Lorentzian case. The relevant 
homotopy groups are respectively $\pi _3(SO(8))=Z$ and $\pi _3(SO(6,2))=\pi _3(SO(6))=Z$.}, 
that implies a quantum statistics for the branes dictated by the coefficient of the action.  
 
The fact that
 $S=c \int _{\mathcal{M}^3} \mathcal{C}_{3}$ is a topological invariant in this case is crucial, because it means that for any off-shell interchange of branes we will get the same phase.

\section{Conclusions}

Studying the model considered in this work one has an odd feeling, like watching a film where the same actor plays different roles, as
it is in a crossroad of many central themes of physics and mathematics, as gauge invariance and characteristic classes, monopoles and instantons, quantum phases and statistics, QHE, Hopf maps, quaternions, perhaps exotic spheres, 
and also a strange mixing of space-time and internal coordinates, related with isomorphisms as $S^3\cong SU(2)$ and their close connection to $SO(3)$, and $SO(4)\cong SU(2)\times SU(2)$,  and the various roles of quaternions in the constructions considered. Many aspects of this model remain to be explored. 

Among them are possible solutions with non zero curvature $F\neq 0$, and the issue 
of small perturbations around different static solutions and their spectrum. 

I do not know whether exotic spheres are euclidean solutions of the theory. I looked for a superbly elegant proof that they are, like the proof that self dual configurations are solutions in the original BPST article \cite{Belavin:1975fg}, and did not find it,  but such proof may well be possible, or maybe some hands on calculation is required.

The generic case of Transgressions, with their two dynamical gauge fields, 
and branes with boundaries (\'a la \cite{Nepomechie1984}) should be researched.

The question of topological quantum phases and statistics deserves further studies. 
Also the gauge group in the euclidean case is $SO(8)$, with the property known a "triality", that when looking at small perturbations may be 
another way to look into this subject.

The apparition of a scale $l$ when the vielbein was introduced could be justified in a most natural way, for instance
if the two sets of gauge fields of the Transgression are considered, one may consider one of them as 
a "ruler" against which to "measure" the other. For instance one may consider a constant curvature solution for $\overline{A}$ with an 
arbitrary scale $l$, 
that "measure" $A$ by comparison\footnote{Like the problem in classic 
Euclidean geometry of determining with straight-edge and compasses the square root of a number, that only makes sense by defining the unit interval, as the straight line has no intrinsic scale.}. 
That may be a natural way to obtain a scale from a theory without one.

The issue of chirality may be investigated, as it could naturally arise in theories that live in the odd dimensional boundary of an even dimensional manifold as Transgressions do. Also, as Yamagishi suggest in the closing paragraph of \cite{Yamagishi:1983dw}, 
solutions with a non zero Pontryagin number should have chiral zero modes among their small perturbations, from the point of view of Index theorems.

\appendix

\section{Review of Quaternions and BPST Instantons}

\subsection{Complex numbers and the Dirac's Monopole in stereographic coordinates}

The calculations in this Subsection can be seen as a simpler version of the ones in the next. 
See for instance Chapter 0 of Ref.\cite{Naber:1997yu}.

The $U(1)$ gauge potential of a magnetic monopole is $A^{\pm}=\frac{g}{4\pi }[-\cos\theta \pm 1]d\phi $, where $A^+ $ is the "northern hemisphere " potential, defined everywhere but in the "south pole"  $\theta =\pi $, 
and $A^-$ is the "southern hemisphere " potential, defined everywhere but in the "north pole"  $\theta = 0$. These potentials differ by a gauge transformation $A^+-A^- =d(\frac{g}{2\pi } \phi )$.

The stereographic projection on a plane tangent to the south pole (thought as the complex plane with coordinates $z=x+i y$), through lines drawn from the northern pole assigning the intersection with the sphere to the intersection with the plane, is given by

\begin{eqnarray}
A^- (z)=-\frac{g}{4\pi }\left[\cos\theta +1\right]d\phi =-\frac{g}{2\pi } Im\left[ \frac{z^*dz}{1+\mid z\mid ^2}\right]\\
A^+ (z)=-\frac{g}{4\pi }\left[\cos\theta -1\right]d\phi = \frac{g}{2\pi } Im\left[ \frac{z^*dz}{\mid z\mid ^2 (1+\mid z\mid ^2)}\right]=A^- (\frac{1}{z})
\end{eqnarray}

Naturally  $A^+$ and $A^-$ also differ by a gauge transformation in this formulation, with gauge group element $\exp [i\frac{g}{2\pi }\phi ]$, 
that is $A^+=A^-+d[\frac{g}{2\pi }\phi ]=\exp [-i\frac{g}{2\pi }\phi ][A^- +d]\exp [i\frac{g}{2\pi }\phi ]$.

The stereographically projected vielbein is

$$E=\frac{2 dz}{1+\mid z\mid ^2 }$$

and the spin connection $\Omega $ corresponding to the zero torsion condition 

$$T=dE+\Omega E=0$$

is 

$$\Omega  =-2 Im\left[ \frac{z^*dz}{1+\mid z\mid ^2}\right] =\frac{4\pi }{g}A^-$$

and its curvature is

$$R=d\Omega =\frac{-2 dz^* dz}{[1+\mid z\mid ^2]^2}=\frac{i}{2}E^* E =Im[E^* E].$$

Notice that this curvature is real, that is $R=R^*$.

\subsection{Quaternions}

For the content of this and the next two review Subsections see Refs.\cite{atiyah1979geometry,Yamagishi:1983dw,Gherardini:2023uyx,Berman:2024yoa,SchettiniGherardini:2025ydb}. Also Ref.\cite{Trautman:1977im} is very relevant to all the Appendix, and it also relates Monopoles and Instantons to Hopf fibrations.

A quaternion can be written as $x=\sum _{a=0}^3x^a e_a$, with $e_0=1$ and the imaginary units $e_1=i$, $e_2=j$ and $e_3=k$. 
The imaginary units satisfy $e_ie_j=-\delta _{ij}+\epsilon _{ijk}e_k$, with $i,j,k=1,2,3$. That implies $e_i^2=-1$ and $e_ie_j=\epsilon _{ijk} e_k$,  if $i\neq j$.

If $x=x^0+x^1e_1+x^2e_2+x^3e_3$ , then its complex conjugate is $\overline{x}=x^0-x^1e_1-x^2e_2-x^3e_3$, or $\overline{1}=1$ and $\overline{e}_i=-e_i$.
The real part is $Re[x]=\frac{x+\overline{x}}{2}=x^0$ and the imaginary part is $Im[x]=\frac{x-\overline{x}}{2}=x^1e_1+x^2e_2+x^3e_3=x^ie_i$. That means $x^a=Re[x\overline{e}^a]$.
The modulus is $\mid x\mid ^2=x\overline{x}=\overline{x}x=(x^0)^2+(x^1)^2+(x^2)^2+(x^3)^2$. The inverse, if $\mid  x\mid \neq 0$, is $x^{-1}=\frac{x}{\mid x\mid ^2}$. 
It also holds that $\overline{xy}=\overline{y}\,\overline{x}$ and that $|xy|=|x|\, |y|$.

The expression $e_{[m}\overline{e}_{n]}\equiv \frac{1}{2}[e_m\overline{e}_n-e_n\overline{e}_m]$ is self-dual, 
that is $e_{[m}\overline{e}_{n]}=\frac{1}{2}\epsilon _{mnpq}e_{[p}\overline{e}_{q]}$, with $m,n,p,q=0,1,2,3$, 
while $\overline{e}_{[m}e_{n]}\equiv \frac{1}{2}[\overline{e}_me_n-\overline{e}_n e_m]$ is anti self-dual $\overline{e}_{[m}e_{n]}=-\frac{1}{2}\epsilon _{mnpq}\overline{e}_{[p}e_{q]}$.

The differential is $dx=\sum _{a=0}^3e_a dx^a$. It holds that $dx d\overline{x}=-^0\eta ^i_{mn} e_idx^mdx^n$, 
with $~^0\eta ^i_{mn}=\epsilon _{imn0}-\delta _{im}\delta _{n0}+\delta _{in}\delta _{m0}$ the self-dual  't Hooft symbol. 
Also $d\overline{x}dx=-^0\overline{\eta } ^i_{mn}e_idx^mdx^n$, with $~^0\overline{\eta } ^i_{mn}=\epsilon _{imn0}+\delta _{mi}\delta _{n0}-\delta _{m0}\delta _{ni}$ the anti self-dual 't Hooft symbol. I am using the convention $\epsilon _{1230}=+1$.

\subsection{BPST Instantons}

The formulation of the BPST $SU(2)$ anti-instanton gauge potential analogous to the formulation of the monopole given in the previous section is given in quaternionic , instead of complex, notation by

\begin{equation}
A=Im\left[ \frac{\overline{x}dx}{1+|x|^2 } \right],
\end{equation}

while the corresponding field strength is

\begin{equation}
F= dA+A^2=Im\left[ \frac{d\overline{x}dx}{[1+|x|^2 ]^2} \right].
\end{equation}

To verify the previous expression the relationship $Im[\overline{x}dx] Im[\overline{x}dx] =Im[(\overline{x}dx)(\overline{x}dx)]$ is required.

The corresponding expressions for a BPST instanton are obtained by intercharging $x \leftrightarrow \overline{x}$.

When $|x|\rightarrow\infty$ then $A\rightarrow Im\left[\frac{\overline{x}dx}{|x|^2}\right]=\left(\frac{x}{|x|}\right)^{-1}d\left(\frac{x}{|x|}\right)$, which is pure gauge and ill-defined, as that limit should correspond to a pole of $S^4$ opposite to the one at $x=0$. This is solved by consider another patch in that hemisphere with variable $y=x^{-1}$, which corresponds to a gauge transformation of the gauge potential. Explicitly

\begin{equation}
A'=Im\left[ \frac{\overline{y}dy}{1+|y|^2 } \right]=Im\left[ \frac{\overline{x}dx}{[1+|x|^2]|x|^2 } \right]
\end{equation}

and 

\begin{equation}
A'=g^{-1}Ag+g^{-1}dg
\end{equation}

with $g=\frac{\overline{x}}{|x|}$. The gauge potentials are identified in the overlap, corresponding to the identification 
$(x,g_0)\leftrightarrow (y, g'_0)\equiv (x^{-1},\left(\frac{x}{|x|}\right)g_0)$ for an arbitrary unit quaternion $g_0$, as a group element of $SU(2)$, defining an $S^3$ bundle over $S^4$. 
Alternatively, we can choose another action of $SU(2)$, defined by $(x,g_0)\leftrightarrow (y, g''_0)\equiv (x^{-1},g_0\left(\frac{x}{|x|}\right)^{-1})$. The right and left actions of $SU(2)$ correspond to the two components of $SO(4)\backsimeq (SU(2)\times SU(2))/Z^2$.
 
The Pontryagin number, an integer topological invariant,  is 

\begin{equation}
n=-\frac{1}{4\pi ^2}\int _{S^4} Re[F^2]=\frac{1}{12\pi ^2}\int _{S^3}Re[g^{-1}dg]
\end{equation}
 which for the transition function $g=\frac{x}{|x|}$, corresponding to $g_0 \mapsto \left(\frac{x}{|x|}\right)g_0$, gives $n=+1$. 
 If instead $g=\left(\frac{x}{|x|}\right)^k$, corresponding to $g_0 \mapsto \left(\frac{x}{|x|}\right)^kg_0$, the Pontryagin number is $n=k$.
 
 \subsection{Vielbein, Spin Connection and Curvature}

If we parametrize $S^3$ with unit quaternions $y$, with $| y |=1$, we have $0=d|y|^2=dy\overline{y}+yd\overline{y}$, or $dy\overline{y}=-yd\overline{y}$. We choose the vielbein in $S^3$ as
$\epsilon\equiv dy\overline{y}=-yd\overline{y}$. This vielbein satisfies the Maurer-Cartan identity $d\epsilon =-dyd\overline{y}=-dy\overline{y}yd\overline{y}=\epsilon ^2$. The metric can be read from the line element $ds^2=Re[\epsilon \otimes \overline \epsilon]$.

The zero torsion condition for $S^3$ reads $d\epsilon +\omega \epsilon +\epsilon\omega =0$, and is satisfied by $\omega =-\frac{\epsilon }{2}$. 
The curvature of $S^3$ is $r=d\omega +\omega ^2=-\frac{\epsilon ^2}{2}+\frac{\epsilon ^2}{4}=-\frac{\epsilon ^2}{4}$.

The vielbein $\epsilon$ is called a left invariant Maurer-Cartan form, while the alternative definition $\epsilon '\equiv \overline{y}dy =-d\overline{y} y$, that satisfies the Maurer-Cartan equation
$d\epsilon '=-(\epsilon ' )^2$, is called a right invariant Maurer-Cartan form.  That is because the $SO(4)\simeq  \frac{SU(2)\times SU(2)}{Z_2}$ isometry of $S^3$ is realized as $y\mapsto uy\overline{v}$, with
$u$ and $v$ unitary quaternions that represent each, left and right,  $SU(2)$ factor.

As for $S^4$, we parametrize it in two patches, for instance to discs $D^4$, each locally diffeomorphic to $R^4\simeq H$, and therefore we will use quaternions $x, x'\in H$ , with $x'=x^{-1}$ identified in the overlap, for instance for $|x|=|x'|=1$. The vielbein is taken as $E=\frac{2 dx}{1+|x|^2}$, and the line element is then $ds^2=Re[E\otimes\overline{E}]$.

The zero torsion condition reads $dE+\Omega _LE+E\Omega _R=0$,which is solved by the left and right spin connections 
$\Omega _L=\frac{Im[xd\overline{x}]}{1+|x|^2}$ and $\Omega _R=\frac{Im[\overline{x}dx]}{1+|x|^2}$, identical to the anti-instanton and instanton potentials. 
The corresponding curvatures are then $R_L=d\Omega _L+\Omega _L^2=\frac{dxd\overline{x}}{[1+|x|^2]^2}$ and $R_R=d\Omega _R+\Omega _R^2=\frac{d\overline{x}dx}{[1+|x|^2]^2}$. 
Notice that $R_L=\frac{1}{4}E\overline{E}$ and $R_R=\frac{1}{4}\overline{E}E$.

\acknowledgments

While working in the present article I received financial support from the {\it Sistema Nacional de Investigadores} (SNI), of the
{\it Agencia Nacional de Investigaci\'on e Innovaci\'on} (ANII) of Uruguay.

\bibliography{Encore-bibfyle}

@book{Weinberg, 
place={New York}, 
title={Dreams of a final theory}, 
publisher={Pantheon Books}, 
author={Weinberg, Steven}, 
year={1992}}

@book{einstein1933method,
  title={On the Method of Theoretical Physics},
  author={Einstein, A.},
  lccn={33020372},
  series={Herbert Spencer lecture},
  url={https://books.google.com.uy/books?id=OSTPAAAAMAAJ},
  year={1933},
  publisher={Oxford University Press}
}

@book{Yang:2005yrp,
    author = "Yang, Chen Ning",
    title = "{Selected Papers (1945{\textendash}1980) of Chen Ning Yang}: {With Commentary}",
    doi = "10.1142/5853",
    isbn = "978-981-256-367-5, 978-981-4480-13-0",
    publisher = "World Scientific",
    address = "Singapore, Singapore",
    series = "World Scientific Series in 20th Century Physics",
    volume = "36",
    year = "2005"
}

@inbook{Zhang_2004, 
place={Cambridge}, 
title={To see a world in a grain of sand}, 
booktitle={Science and Ultimate Reality: Quantum Theory, Cosmology, and Complexity}, 
publisher={Cambridge University Press}, 
author={Zhang, Shou-Cheng}, 
editor={Barrow, John D. and Davies, Paul C. W. and Harper, Jr, Charles L.Editors}, 
year={2004}, 
pages={667–690}}

@book{Wen:2004ym,
  title={Quantum Field Theory of Many-Body Systems: From the Origin of Sound to an Origin of Light and Electrons},
  author={Wen, X.G.},
  isbn={9780198530947},
  lccn={2004301677},
  series={Oxford Graduate Texts},
  url={https://books.google.com.uy/books?id=RYESDAAAQBAJ},
  year={2004},
  publisher={OUP Oxford}
}

@article{Wilczek:1981du,
    author = "Wilczek, Frank",
    title = "{Magnetic Flux, Angular Momentum, and Statistics}",
    reportNumber = "NSF-ITP-81-117",
    doi = "10.1103/PhysRevLett.48.1144",
    journal = "Phys. Rev. Lett.",
    volume = "48",
    pages = "1144--1146",
    year = "1982"
}

@article{Wilczek:1982wy,
    author = "Wilczek, Frank",
    title = "{Quantum Mechanics of Fractional Spin Particles}",
    reportNumber = "NSF-ITP-82-56",
    doi = "10.1103/PhysRevLett.49.957",
    journal = "Phys. Rev. Lett.",
    volume = "49",
    pages = "957--959",
    year = "1982"
}

@article{Wilczek:1983cy,
    author = "Wilczek, Frank and Zee, A.",
    title = "{Linking Numbers, Spin, and Statistics of Solitons}",
    reportNumber = "NSF-ITP-83-148",
    doi = "10.1103/PhysRevLett.51.2250",
    journal = "Phys. Rev. Lett.",
    volume = "51",
    pages = "2250--2252",
    year = "1983"
}

@article{Wu:1984kd,
    author = "Wu, Yong-Shi and Zee, A.",
    title = "{Comments on the Hopf Lagrangian and Fractional Statistics of Solitons}",
    reportNumber = "DOE-ER-40048-18 P4",
    doi = "10.1016/0370-2693(84)90126-6",
    journal = "Phys. Lett. B",
    volume = "147",
    pages = "325--329",
    year = "1984"
}

@incollection{Nepomechie1984, 
place={Bristol}, 
booktitle={Quantum Field Theory and Quantum Statistics. Essays in Honour of the Sixtieth Birthday of E. S. Fradkin.}, 
publisher={Hilger}, 
author={Nepomechie, Rafael and Zee, Anthony},
editor={Batalin, I. and Isham, C. and Vilkovisky, G.},
title={Phase interaction between extended objects}, 
pages={467-472},
year={1987}}

@article{Wu:1988py,
    author = "Wu, Y. S. and Zee, A.",
    title = "{Membranes, Higher Hopf Maps, and Phase Interactions}",
    doi = "10.1016/0370-2693(88)90882-9",
    journal = "Phys. Lett. B",
    volume = "207",
    pages = "39--43",
    year = "1988"
}

@article{Tze:1989jv,
    author = "Tze, Hsiung Chia and Nam, Soonkeon",
    title = "{Topological Phase Entanglements of Membrane Solitons in Division Algebra $\sigma$ Models With a Hopf Term}",
    reportNumber = "VPI-IHEP-88-10",
    doi = "10.1016/0003-4916(89)90005-5",
    journal = "Annals Phys.",
    volume = "193",
    pages = "419--471",
    year = "1989"
}

@article{Jain:1989tx,
    author = "Jain, J. K.",
    title = "{Composite fermion approach for the fractional quantum Hall effect}",
    doi = "10.1103/PhysRevLett.63.199",
    journal = "Phys. Rev. Lett.",
    volume = "63",
    pages = "199--202",
    year = "1989"
}

@article{Haldane:1983xm,
    author = "Haldane, F. D. M.",
    title = "{Fractional quantization of the Hall effect: A Hierarchy of incompressible quantum fluid states}",
    doi = "10.1103/PhysRevLett.51.605",
    journal = "Phys. Rev. Lett.",
    volume = "51",
    pages = "605--608",
    year = "1983"
}

@article{Zhang:1988wy,
    author = "Zhang, S. C. and Hansson, T. H. and Kivelson, S.",
    title = "{An effective field theory model for the fractional quantum hall effect}",
    reportNumber = "NSF-ITP-88-119",
    doi = "10.1103/PhysRevLett.62.82",
    journal = "Phys. Rev. Lett.",
    volume = "62",
    pages = "82--85",
    year = "1988"
}

@article{Zhang:1992eu,
    author = "Zhang, Shou-Cheng",
    title = "{The Chern-Simons-Landau-Ginzburg theory of the fractional quantum Hall effect}",
    doi = "10.1142/S0217979292000037",
    journal = "Int. J. Mod. Phys. B",
    volume = "6",
    pages = "25--58",
    year = "1992"
}

@article{Frohlich:1991wb,
    author = "Frohlich, J. and Zee, A.",
    title = "{Large scale physics of the quantum Hall fluid}",
    reportNumber = "NSF-ITP-91-31",
    doi = "10.1016/0550-3213(91)90275-3",
    journal = "Nucl. Phys. B",
    volume = "364",
    pages = "517--540",
    year = "1991"
}

@article{Zee:1995avy,
    author = "Zee, A.",
    editor = "Geyer, H. B.",
    title = "{Quantum Hall fluids}",
    eprint = "cond-mat/9501022",
    archivePrefix = "arXiv",
    doi = "10.1007/BFb0113369",
    journal = "Lect. Notes Phys.",
    volume = "456",
    pages = "99--153",
    year = "1995"
}

@article{Zhang:2001xs,
    author = "Zhang, Shou-Cheng and Hu, Jiang-ping",
    title = "{A Four-dimensional generalization of the quantum Hall effect}",
    eprint = "cond-mat/0110572",
    archivePrefix = "arXiv",
    doi = "10.1126/science.294.5543.823",
    journal = "Science",
    volume = "294",
    pages = "823",
    year = "2001"
}

@article{Bernevig:2002eq,
    author = "Bernevig, B. Andrei and Chern, Chyh-Hong and Hu, Jiang-Ping and Toumbas, Nicolaos and Zhang, Shou-Cheng",
    title = "{Effective field theory description of the higher dimensional quantum Hall liquid}",
    eprint = "cond-mat/0206164",
    archivePrefix = "arXiv",
    doi = "10.1006/aphy.2002.6292",
    journal = "Annals Phys.",
    volume = "300",
    pages = "185",
    year = "2002"
}

@article{Bernevig:2003yz,
    author = "Bernevig, Bogdan A. and Hu, Jiang-ping and Toumbas, Nicolaos and Zhang, Shou-Cheng",
    title = "{The Eight-dimensional quantum Hall effect and the octonions}",
    eprint = "cond-mat/0306045",
    archivePrefix = "arXiv",
    doi = "10.1103/PhysRevLett.91.236803",
    journal = "Phys. Rev. Lett.",
    volume = "91",
    pages = "236803",
    year = "2003"
}

@article{Karabali:2002im,
    author = "Karabali, Dimitra and Nair, V. P.",
    title = "{Quantum Hall effect in higher dimensions}",
    eprint = "hep-th/0203264",
    archivePrefix = "arXiv",
    reportNumber = "CCNY-HEP-02-03",
    doi = "10.1016/S0550-3213(02)00634-X",
    journal = "Nucl. Phys. B",
    volume = "641",
    pages = "533--546",
    year = "2002"
}

@article{Karabali:2003bt,
    author = "Karabali, Dimitra and Nair, V. P.",
    title = "{The effective action for edge states in higher dimensional quantum Hall systems}",
    eprint = "hep-th/0307281",
    archivePrefix = "arXiv",
    reportNumber = "CCNY-HEP-03-4",
    doi = "10.1016/j.nuclphysb.2003.11.020",
    journal = "Nucl. Phys. B",
    volume = "679",
    pages = "427--446",
    year = "2004"
}

@article{Karabali:2004km,
    author = "Karabali, Dimitra and Nair, V. P.",
    title = "{Edge states for quantum Hall droplets in higher dimensions and a generalized WZW model}",
    eprint = "hep-th/0403111",
    archivePrefix = "arXiv",
    reportNumber = "CCNY-HEP-04-2",
    doi = "10.1016/j.nuclphysb.2004.07.014",
    journal = "Nucl. Phys. B",
    volume = "697",
    pages = "513--540",
    year = "2004"
}

@article{Agarwal:2024fgp,
    author = "Agarwal, Abhishek and Karabali, Dimitra and Nair, V. P.",
    title = "{Fractional quantum Hall effect in higher dimensions}",
    eprint = "2410.14036",
    archivePrefix = "arXiv",
    primaryClass = "hep-th",
    doi = "10.1103/PhysRevD.111.025002",
    journal = "Phys. Rev. D",
    volume = "111",
    number = "2",
    pages = "025002",
    year = "2025"
}

@article{Karabali:2026non,
    author = "Karabali, Dimitra and Nair, V. P.",
    title = "{Higher dimensional quantum Hall effect and the analog of W{\ensuremath{\infty}}-algebra}",
    eprint = "2606.28219",
    archivePrefix = "arXiv",
    primaryClass = "hep-th",
    doi = "10.1103/l8nj-kpgg",
    journal = "Phys. Rev. D",
    volume = "114",
    number = "6",
    pages = "065025",
    year = "2026"
}

@article{Hasebe:2016tjg,
    author = "Hasebe, Kazuki",
    title = "{Higher (Odd) Dimensional Quantum Hall Effect and Extended Dimensional Hierarchy}",
    eprint = "1612.05853",
    archivePrefix = "arXiv",
    primaryClass = "hep-th",
    doi = "10.1016/j.nuclphysb.2017.03.017",
    journal = "Nucl. Phys. B",
    volume = "920",
    pages = "475--520",
    year = "2017"
}

@article{Hasebe:2014nia,
    author = "Hasebe, Kazuki",
    title = "{Higher Dimensional Quantum Hall Effect as A-Class Topological Insulator}",
    eprint = "1403.5066",
    archivePrefix = "arXiv",
    primaryClass = "hep-th",
    doi = "10.1016/j.nuclphysb.2014.07.011",
    journal = "Nucl. Phys. B",
    volume = "886",
    pages = "952--1002",
    year = "2014"
}

@article{Belavin:1975fg,
    author = "Belavin, A. A. and Polyakov, Alexander M. and Schwartz, A. S. and Tyupkin, Yu. S.",
    editor = "Taylor, J. C.",
    title = "{Pseudoparticle Solutions of the Yang-Mills Equations}",
    doi = "10.1016/0370-2693(75)90163-X",
    journal = "Phys. Lett. B",
    volume = "59",
    pages = "85--87",
    year = "1975"
}

@book{atiyah1979geometry,
  title={Geometry of Yang-Mills Fields},
  author={Atiyah, M.F. and Accademia nazionale dei Lincei and Scuola normale superiore (Italy)},
  isbn={9788876423031},
  series={Fermi lectures},
  url={https://books.google.com.uy/books?id=VcjvAAAAMAAJ},
  year={1979},
  publisher={Scuola Normale Superiore}
}

@article{Atiyah:1978ri,
    author = "Atiyah, M. F. and Hitchin, Nigel J. and Drinfeld, V. G. and Manin, Yu. I.",
    editor = "Shifman, Mikhail A.",
    title = "{Construction of Instantons}",
    doi = "10.1016/0375-9601(78)90141-X",
    journal = "Phys. Lett. A",
    volume = "65",
    pages = "185--187",
    year = "1978"
}

@article{Trautman:1977im,
    author = "Trautman, Andrzej",
    title = "{Solutions of the Maxwell and Yang-Mills Equations Associated with Hopf Fibrings}",
    reportNumber = "ITP-SB-77-24",
    doi = "10.1007/BF01811088",
    journal = "Int. J. Theor. Phys.",
    volume = "16",
    pages = "561",
    year = "1977"
}

@article{Dirac:1931kp,
    author = "Dirac, Paul Adrien Maurice",
    title = "{Quantised singularities in the electromagnetic field,}",
    reportNumber = "RX-722",
    doi = "10.1098/rspa.1931.0130",
    journal = "Proc. Roy. Soc. Lond. A",
    volume = "133",
    number = "821",
    pages = "60--72",
    year = "1931"
}

@article{Dirac:1948um,
    author = "Dirac, Paul A. M.",
    title = "{The Theory of magnetic poles}",
    doi = "10.1103/PhysRev.74.817",
    journal = "Phys. Rev.",
    volume = "74",
    pages = "817--830",
    year = "1948"
}

@article{Wu:1975es,
    author = "Wu, Tai Tsun and Yang, Chen Ning",
    editor = "Hsu, Jong-Ping and Fine, D.",
    title = "{Concept of Nonintegrable Phase Factors and Global Formulation of Gauge Fields}",
    reportNumber = "ITP-SB-75-31",
    doi = "10.1103/PhysRevD.12.3845",
    journal = "Phys. Rev. D",
    volume = "12",
    pages = "3845--3857",
    year = "1975"
}

@article{Hull:2024uwz,
    author = "Hull, C. M.",
    title = "{Monopoles, Dirac strings and generalised symmetries}",
    eprint = "2411.18741",
    archivePrefix = "arXiv",
    primaryClass = "hep-th",
    reportNumber = "Imperial-TP-2024-CH-7",
    doi = "10.1007/JHEP09(2025)019",
    journal = "JHEP",
    volume = "09",
    pages = "019",
    year = "2025"
}

@article{Hull:2025yww,
    author = "Hull, Chris",
    title = "{Coupling self-dual p-form Gauge fields to self-dual branes}",
    eprint = "2501.10566",
    archivePrefix = "arXiv",
    primaryClass = "hep-th",
    reportNumber = "Imperial-TP-2025-CH-01",
    doi = "10.1088/1751-8121/adbf76",
    journal = "J. Phys. A",
    volume = "58",
    number = "12",
    pages = "125202",
    year = "2025"
}

@book{Naber:1997yu,
    author = "Naber, G. L.",
    title = "{Topology, geometry, and gauge fields: Foundations}",
    doi = "10.1007/978-1-4419-7254-5",
    publisher = "Springer",
    address = "New York, USA",
    year = "1997"
}

@article{Banados:1997df,
    author = "Banados, Maximo",
    title = "{Constant curvature black holes}",
    eprint = "gr-qc/9703040",
    archivePrefix = "arXiv",
    doi = "10.1103/PhysRevD.57.1068",
    journal = "Phys. Rev. D",
    volume = "57",
    pages = "1068--1072",
    year = "1998"
}

@article{Guilleminot:2017hfy,
    author = "Guilleminot, Pablo and Olea, Rodrigo and Petrov, Alexander N.",
    title = "{Constant curvature black holes in Einstein AdS gravity: Euclidean action and thermodynamics}",
    eprint = "1712.09055",
    archivePrefix = "arXiv",
    primaryClass = "gr-qc",
    doi = "10.1103/PhysRevD.97.064046",
    journal = "Phys. Rev. D",
    volume = "97",
    number = "6",
    pages = "064046",
    year = "2018"
}

@article{Milnor,
    author = "Milnor, John",
    title = "{On Manifolds Homeomorphic to the 7-Sphere}",
    doi = "10.2307/1969983",
    journal = "Annals of Mathematics",
    volume = "64",
    pages = "399--405",
    year = "1956"
}

@article{Yamagishi:1983dw,
    author = "Yamagishi, Kengo",
    title = "{SUPERGRAVITY ON SEVEN-DIMENSIONAL HOMOTOPY SPHERES}",
    reportNumber = "OU-HET-55",
    doi = "10.1016/0370-2693(84)90981-X",
    journal = "Phys. Lett. B",
    volume = "134",
    pages = "47--50",
    year = "1984"
}

@article{Freund:1984am,
    author = "Freund, Peter G. O.",
    title = "{HIGHER DIMENSIONAL UNIFICATION}",
    reportNumber = "EFI-84-7-CHICAGO",
    journal = "Physica D",
    volume = "15",
    pages = "263",
    year = "1985"
}

@article{Witten:1985xe,
    author = "Witten, Edward",
    editor = "Salam, A. and Sezgin, E.",
    title = "{GLOBAL GRAVITATIONAL ANOMALIES}",
    reportNumber = "PRINT-85-0246 (PRINCETON)",
    doi = "10.1007/BF01212448",
    journal = "Commun. Math. Phys.",
    volume = "100",
    pages = "197",
    year = "1985"
}

@article{Brans:1992mj,
    author = "Brans, Carl H. and Randall, Duane",
    title = "{Exotic differentiable structures and general relativity}",
    eprint = "gr-qc/9212003",
    archivePrefix = "arXiv",
    reportNumber = "LOYOLA-B9202",
    doi = "10.1007/BF00758828",
    journal = "Gen. Rel. Grav.",
    volume = "25",
    pages = "205",
    year = "1993"
}

@article{Brans:1994hq,
    author = "Brans, Carl H.",
    title = "{Exotic smoothness and physics}",
    eprint = "gr-qc/9405010",
    archivePrefix = "arXiv",
    doi = "10.1063/1.530761",
    journal = "J. Math. Phys.",
    volume = "35",
    pages = "5494--5506",
    year = "1994"
}

@article{Brans:1994wv,
    author = "Brans, Carl H.",
    title = "{Localized exotic smoothness}",
    eprint = "gr-qc/9404003",
    archivePrefix = "arXiv",
    reportNumber = "IASSNS-HEP-94-22",
    doi = "10.1088/0264-9381/11/7/015",
    journal = "Class. Quant. Grav.",
    volume = "11",
    pages = "1785--1792",
    year = "1994"
}

@article{Brans:1996ix,
    author = "Brans, Carl H.",
    editor = "Cho, Y. M. and Lee, C. H. and Kim, S. W.",
    title = "{Exotic smoothness on space-time}",
    eprint = "gr-qc/9604048",
    archivePrefix = "arXiv",
    doi = "10.1007/978-94-011-5812-1_27",
    journal = "Astrophys. Space Sci. Libr.",
    volume = "211",
    pages = "160",
    year = "1997"
}

@article{Asselmeyer:1996bh,
    author = "Asselmeyer, T.",
    title = "{Generation of source terms in general relativity by differential structures}",
    eprint = "gr-qc/9610009",
    archivePrefix = "arXiv",
    doi = "10.1088/0264-9381/14/3/016",
    journal = "Class. Quant. Grav.",
    volume = "14",
    pages = "749--758",
    year = "1997"
}

@article{Asselmeyer-Maluga:2015qka,
    author = "Asselmeyer-Maluga, Torsten and Brans, Carl H.",
    title = "{How to include fermions into General relativity by exotic smoothness}",
    eprint = "1502.02087",
    archivePrefix = "arXiv",
    primaryClass = "gr-qc",
    doi = "10.1007/s10714-015-1872-x",
    journal = "Gen. Rel. Grav.",
    volume = "47",
    number = "3",
    pages = "30",
    year = "2015"
}

@article{Gherardini:2023uyx,
    author = "Gherardini, T. Schettini",
    title = "{Exotic spheres{\textquoteright} metrics and solutions via Kaluza-Klein techniques}",
    eprint = "2309.01703",
    archivePrefix = "arXiv",
    primaryClass = "hep-th",
    doi = "10.1007/JHEP12(2023)100",
    journal = "JHEP",
    volume = "12",
    pages = "100",
    year = "2023"
}

@article{Berman:2024yoa,
    author = "Berman, David S. and Cederwall, Martin and Gherardini, Tancredi Schettini",
    title = "{Curvature of an exotic 7-sphere}",
    eprint = "2410.01909",
    archivePrefix = "arXiv",
    primaryClass = "hep-th",
    reportNumber = "QMUL-PH-24-22",
    doi = "10.1016/j.geomphys.2025.105590",
    journal = "J. Geom. Phys.",
    volume = "216",
    pages = "105590",
    year = "2025"
}

@phdthesis{SchettiniGherardini:2025ydb,
    author = "Schettini Gherardini, Tancredi",
    title = "{A Physicist{\textquoteright}s Visit to Exotic Spheres}",
    eprint = "2604.21997",
    archivePrefix = "arXiv",
    primaryClass = "hep-th",
    school = "Queen Mary, U. of London (main)",
    year = "2025"
}

@inproceedings{Zee:1989fr,
    author = "Zee, A.",
    title = "{Musings on quantum statistics}",
    booktitle = "{Fundamental Aspects of Quantum Theory: Conference on Foundations of Quantum Mechanics to Celebrate 30 Years of the Aharonov-Bohm Effect (AB Conference)}",
    month = "12",
    year = "1989"
}

@inproceedings{Zee:1992va,
    author = "Zee, A.",
    title = "{Quantum phase in action}",
    booktitle = "{International Conference on Fundamental Aspects of Quantum Theory to Celebrate the 60th Birthday of Yakir Aharonov}",
    pages = "109--113",
    month = "12",
    year = "1992"
}

@article{Jackiw:1976xx,
    author = "Jackiw, R. and Rebbi, C.",
    title = "{Spin from Isospin in a Gauge Theory}",
    reportNumber = "MIT-CTP-524",
    doi = "10.1103/PhysRevLett.36.1116",
    journal = "Phys. Rev. Lett.",
    volume = "36",
    pages = "1116",
    year = "1976"
}

@article{Jackiw:1975fn,
    author = "Jackiw, R. and Rebbi, C.",
    title = "{Solitons with Fermion Number 1/2}",
    reportNumber = "MIT-CTP-517",
    doi = "10.1103/PhysRevD.13.3398",
    journal = "Phys. Rev. D",
    volume = "13",
    pages = "3398--3409",
    year = "1976"
}

@article{Hasenfratz:1976gr,
    author = "Hasenfratz, P. and 't Hooft, Gerard",
    title = "{A Fermion-Boson Puzzle in a Gauge Theory}",
    reportNumber = "Print-76-0193 (UTRECHT)",
    doi = "10.1103/PhysRevLett.36.1119",
    journal = "Phys. Rev. Lett.",
    volume = "36",
    pages = "1119",
    year = "1976"
}

@incollection{Yang, 
place={Cambridge}, 
booktitle={Schroedinger: Centenary Celebration of a Polymath}, 
publisher={Cambridge University Press}, 
author={Yang, Chen-Ning},
editor={Kilmister, C. W.},
title={Square root of minus one, complex phases and Erwin Schroedinger}, 
pages={55-64},
year={1987}}

@inproceedings{Yang:2007upv,
    author = "Yang, Chen Ning",
    title = "{Thematic Melodies of Twentieth Century Theoretical Physics: Quantization, Symmetry and Phase Factor}",
    booktitle = "{3rd Asia Pacific Conference on Few-Body Problems in Physics (APFB05)}",
    doi = "10.1142/9789812706881_0001",
    pages = "1--8",
    year = "2007"
}

@article{Yang:2012bim,
    author = "Yang, Chen Ning",
    title = "{Quantum numbers, Chern classes, and a bodhisattva}",
    doi = "10.1063/pt.3.1398",
    journal = "Phys. Today",
    volume = "65",
    number = "1",
    pages = "33--34",
    year = "2012"
}

@article{Montero:2025ayi,
    author = "Montero, Miguel and Zapata, Luis",
    title = "{M-theory boundaries beyond supersymmetry}",
    eprint = "2504.06985",
    archivePrefix = "arXiv",
    primaryClass = "hep-th",
    reportNumber = "IFT-25-033",
    doi = "10.1007/JHEP07(2025)090",
    journal = "JHEP",
    volume = "07",
    pages = "090",
    year = "2025"
}

@article{Mora-Nishino20001,
  title   = "{Fundamental extended objects for Chern–Simons supergravity}",
  journal = "Physics Letters B ",
  volume  = "482",
  pages   = "222-232",
  year    = "2000",
  doi     = "10.1016/S0370-2693(00)00535-9",
  archivePrefix = {arXiv},
   eprint = {hep-th/0002077},
  author  = "P. Mora and H. Nishino"
}

@article{Mora:2021wmz,
    author = "Mora, Pablo",
    title = "{Actions, equations of motion and boundary conditions for Chern-Simons branes}",
    eprint = "2105.08863",
    archivePrefix = "arXiv",
    primaryClass = "hep-th",
    doi = "10.1016/j.physletb.2021.136428",
    journal = "Phys. Lett. B",
    volume = "819",
    pages = "136428",
    year = "2021"
}

@article{Mora:2023ngp,
    author = "Mora, Pablo",
    title = "{Chern-Simons branes with enhanced gauge invariance}",
    eprint = "2305.06186",
    archivePrefix = "arXiv",
    primaryClass = "hep-th",
    doi = "10.1007/JHEP07(2023)107",
    journal = "JHEP",
    volume = "07",
    pages = "107",
    year = "2023"
}

@article{Zumino:1983rz,
    author = "Zumino, Bruno and Wu, Yong-Shi and Zee, A.",
    title = "{Chiral Anomalies, Higher Dimensions, and Differential Geometry}",
    reportNumber = "LBL-16443, DOE-ER-40048-18-P3",
    doi = "10.1016/0550-3213(84)90259-1",
    journal = "Nucl. Phys. B",
    volume = "239",
    pages = "477--507",
    year = "1984"
}

@article{Alvarez-Gaume:1984zlq,
    author = "Alvarez-Gaume, Luis and Ginsparg, Paul H.",
    editor = "Salam, A. and Sezgin, E.",
    title = "{The Structure of Gauge and Gravitational Anomalies}",
    reportNumber = "HUTP-84/A016",
    doi = "10.1016/0003-4916(85)90087-9",
    journal = "Annals Phys.",
    volume = "161",
    pages = "423",
    year = "1985",
    note = "[Erratum: Annals Phys. 171, 233 (1986)]"
}

@book{Zee:2003mt,
    author = "Zee, A.",
    title = "{Quantum field theory in a nutshell (Second Ed.)}",
    publisher={Princeton University Press},
    isbn = "978-0-691-14034-6",
    year = "2010"
}

@article{Mora20141,
title = "{Action Principles for Transgression and Chern-Simons AdS Gravities}",
journal = "Journal of High Energy Physics",
volume = "11",
pages = "128",
year = "2014",
archivePrefix = {arXiv},
eprint = {1407.6032},
primaryClass = {hep-th},
doi = "10.1007/JHEP11(2014)128",
author = "P. Mora"
}

@article{MOTZ-2004,
title = "{Finite action principle for Chern-Simons AdS gravity}",
journal = "Journal of High Energy Physics",
volume = "06",
pages = "036",
year = "2004",
archivePrefix = "arXiv",
eprint = "hep-th/0405267",
reportNumber = "CECS-PHY-04-07",
doi = "10.1088/1126-6708/2004/06/036",
author = "Mora, P. and Olea, R. and Troncoso, R. and Zanelli, J."
}

@article{MOTZ-2006,
title = "{Transgression forms and extensions of Chern-Simons gauge theories}",
 journal = "Journal of High Energy Physics",
volume = "02",
pages = "067",
year = "2006",   
archivePrefix = "arXiv",    
eprint = "hep-th/0601081",
reportNumber = "CECS-PHY-06-01",
doi = "10.1088/1126-6708/2006/02/067",
author = "Mora, Pablo and Olea, Rodrigo and Troncoso, Ricardo and Zanelli, Jorge"
}

@article{Cai:2024tyv,
    author = "Cai, Rong-Gen and Li, Li and Zhao, Jun-Kun",
    title = "{Transgression Completion for Chern-Simons Black Hole Thermodynamics}",
    eprint = "2410.00717",
    archivePrefix = "arXiv",
    primaryClass = "gr-qc",
    month = "10",
    year = "2024"
}

@inproceedings{Edelstein:2008ry,
    author = "Edelstein, Jose D. and Zanelli, Jorge",
    title = "{Sources for Chern-Simons theories}",
    booktitle = "{workshop on Quantum Mechanics of Fundamental Systems: the Quest for Beauty and Simplicity: Dedicated to Claudio Bunster on the occasion of his 60th birthday}",
    eprint = "0807.4217",
    archivePrefix = "arXiv",
    primaryClass = "hep-th",
    doi = "10.1007/978-0-387-87499-9_8",
    pages = "107--124",
    year = "2009"
}

@article{Miskovic:2009dd,
    author = "Miskovic, Olivera and Zanelli, Jorge",
    title = "{Couplings between Chern-Simons gravities and 2p-branes}",
    eprint = "0901.0737",
    archivePrefix = "arXiv",
    primaryClass = "hep-th",
    doi = "10.1103/PhysRevD.80.044003",
    journal = "Phys. Rev. D",
    volume = "80",
    pages = "044003",
    year = "2009"
}

@article{Edelstein:2010sx,
    author = "Edelstein, Jose D. and Garbarz, Alan and Miskovic, Olivera and Zanelli, Jorge",
    title = "{Stable p-branes in Chern-Simons AdS supergravities}",
    eprint = "1006.3753",
    archivePrefix = "arXiv",
    primaryClass = "hep-th",
    reportNumber = "AEI-2010-117",
    doi = "10.1103/PhysRevD.82.044053",
    journal = "Phys. Rev. D",
    volume = "82",
    pages = "044053",
    year = "2010"
}

@article{Edelstein:2010sh,
    author = "Edelstein, Jose D. and Garbarz, Alan and Miskovic, Olivera and Zanelli, Jorge",
    editor = "Calzetta, Esteban and Domenech, Graciela and Ferraro, Rafael and Giribet, Gaston and Laura, Roberto and Nunez, Carmen",
    title = "{Naked Singularities, Topological Defects and Brane Couplings}",
    eprint = "1009.4418",
    archivePrefix = "arXiv",
    primaryClass = "hep-th",
    doi = "10.1142/S0218271811019177",
    journal = "Int. J. Mod. Phys. D",
    volume = "20",
    pages = "839--849",
    year = "2011"
}

@article{Edelstein:2011vu,
    author = "Edelstein, Jose D. and Garbarz, Alan and Miskovic, Olivera and Zanelli, Jorge",
    title = "{Geometry and stability of spinning branes in AdS gravity}",
    eprint = "1108.3523",
    archivePrefix = "arXiv",
    primaryClass = "hep-th",
    doi = "10.1103/PhysRevD.84.104046",
    journal = "Phys. Rev. D",
    volume = "84",
    pages = "104046",
    year = "2011"
}

@article{Eguchi:1980jx,
    author = "Eguchi, Tohru and Gilkey, Peter B. and Hanson, Andrew J.",
    title = "{Gravitation, Gauge Theories and Differential Geometry}",
    reportNumber = "PRINT-80-0182 (LBL,BERKELEY)",
    doi = "10.1016/0370-1573(80)90130-1",
    journal = "Phys. Rept.",
    volume = "66",
    pages = "213",
    year = "1980"
}

@article{Green:1989nn,
    author = "Green, Michael B.",
    title = "{Supertranslations, Superstrings and \{Chern-Simons\} Forms}",
    reportNumber = "QMC-89-5",
    doi = "10.1016/0370-2693(89)90233-5",
    journal = "Phys. Lett. B",
    volume = "223",
    pages = "157--164",
    year = "1989"
}

@article{Dixon:1991xz,
    author = "Dixon, John A. and Duff, M.J. and Sezgin, E.",
    title = "{The Coupling of Yang-Mills to extended objects}",
    eprint = "hep-th/9201019",
    archivePrefix = "arXiv",
    reportNumber = "CTP-TAMU-101-91",
    doi = "10.1016/0370-2693(92)90391-G",
    journal = "Phys. Lett. B",
    volume = "279",
    pages = "265--271",
    year = "1992"
}

@article{Achucarro:1987vz,
    author = "Achucarro, A. and Townsend, P.K.",
    editor = "Salam, A. and Sezgin, E.",
    title = "{A Chern-Simons Action for Three-Dimensional anti-De Sitter Supergravity Theories}",
    reportNumber = "Print-87-0078 (CAMBRIDGE)",
    doi = "10.1016/0370-2693(86)90140-1",
    journal = "Phys. Lett. B",
    volume = "180",
    pages = "89",
    year = "1986"
}

@article{Chamseddine:1990gk,
    author = "Chamseddine, Ali H.",
    title = "{Topological gravity and supergravity in various dimensions}",
    reportNumber = "PRINT-90-0350 (ZURICH)",
    doi = "10.1016/0550-3213(90)90245-9",
    journal = "Nucl. Phys. B",
    volume = "346",
    pages = "213--234",
    year = "1990"
}

@article{Chamseddine:1989nu,
    author = "Chamseddine, Ali H.",
    title = "{Topological Gauge Theory of Gravity in Five-dimensions and All Odd Dimensions}",
    reportNumber = "Print-89-0828 (ZURICH)",
    doi = "10.1016/0370-2693(89)91312-9",
    journal = "Phys. Lett. B",
    volume = "233",
    pages = "291--294",
    year = "1989"
}

@article{Zanelli:2012px,
    author = "Zanelli, Jorge",
    title = "{Chern-Simons Forms in Gravitation Theories}",
    eprint = "1208.3353",
    archivePrefix = "arXiv",
    primaryClass = "hep-th",
    doi = "10.1088/0264-9381/29/13/133001",
    journal = "Class. Quant. Grav.",
    volume = "29",
    pages = "133001",
    year = "2012"
}

@article{MuellerHoissen:1990vf,
    author = "Mueller-Hoissen, Folkert",
    title = "{From Chern-Simons to Gauss-Bonnet}",
    reportNumber = "GOET-TP-11/90",
    doi = "10.1016/0550-3213(90)90246-A",
    journal = "Nucl. Phys. B",
    volume = "346",
    pages = "235--252",
    year = "1990"
}

@article{Banados:1996hi,
    author = "Banados, Maximo and Troncoso, Ricardo and Zanelli, Jorge",
    title = "{Higher dimensional Chern-Simons supergravity}",
    eprint = "gr-qc/9601003",
    archivePrefix = "arXiv",
    doi = "10.1103/PhysRevD.54.2605",
    journal = "Phys. Rev. D",
    volume = "54",
    pages = "2605--2611",
    year = "1996"
}

@article{Banados:1993ur,
    author = "Banados, Maximo and Teitelboim, Claudio and Zanelli, Jorge",
    title = "{Dimensionally continued black holes}",
    eprint = "gr-qc/9307033",
    archivePrefix = "arXiv",
    reportNumber = "IASSNS-AST-93-45",
    doi = "10.1103/PhysRevD.49.975",
    journal = "Phys. Rev. D",
    volume = "49",
    pages = "975--986",
    year = "1994"
}

@article{Troncoso1997va,
    author = "Troncoso, Ricardo and Zanelli, Jorge",
    title = "{New gauge supergravity in seven-dimensions and eleven-dimensions}",
    eprint = "hep-th/9710180",
    archivePrefix = "arXiv",
    doi = "10.1103/PhysRevD.58.101703",
    journal = "Phys. Rev. D",
    volume = "58",
    pages = "101703",
    year = "1998"
}

@article{Troncoso1998ng,
    author = "Troncoso, Ricardo and Zanelli, Jorge",
    editor = "Castagnino, M. A. and Nunez, Carmen A.",
    title = "{Gauge supergravities for all odd dimensions}",
    eprint = "hep-th/9807029",
    archivePrefix = "arXiv",
    doi = "10.1023/A:1026614631617",
    journal = "Int. J. Theor. Phys.",
    volume = "38",
    pages = "1181--1206",
    year = "1999"
}

@article{Horava:1997dd,
    author = "Horava, Petr",
    title = "{M theory as a holographic field theory}",
    eprint = "hep-th/9712130",
    archivePrefix = "arXiv",
    reportNumber = "CALT-68-2152",
    doi = "10.1103/PhysRevD.59.046004",
    journal = "Phys. Rev. D",
    volume = "59",
    pages = "046004",
    year = "1999"
}

@article{Ertem:2012qv,
    author = {Ertem, \"Umit and Acik, \"Ozg\"ur},
    title = "{Couplings of gravitational currents with Chern-Simons gravities}",
    eprint = "1211.3289",
    archivePrefix = "arXiv",
    primaryClass = "gr-qc",
    doi = "10.1103/PhysRevD.87.044052",
    journal = "Phys. Rev. D",
    volume = "87",
    number = "4",
    pages = "044052",
    year = "2013"
}

@article{Frey:2019fqz,
    author = "Frey, Andrew R",
    title = "{Dirac branes for Dirichlet branes: Supergravity actions}",
    eprint = "1907.12755",
    archivePrefix = "arXiv",
    primaryClass = "hep-th",
    doi = "10.1103/PhysRevD.102.046017",
    journal = "Phys. Rev. D",
    volume = "102",
    number = "4",
    pages = "046017",
    year = "2020"
}

@article{Kastikainen:2020auf,
    author = "Kastikainen, Jani",
    title = "{Conical defects and holography in topological AdS gravity}",
    eprint = "2006.02803",
    archivePrefix = "arXiv",
    primaryClass = "hep-th",
    doi = "10.1088/1361-6382/abac44",
    journal = "Class. Quant. Grav.",
    volume = "37",
    number = "19",
    pages = "195010",
    year = "2020"
}

\end{document}